\documentclass[preprint,3p, times]{elsarticle}
\pdfoutput=1
\usepackage{amsfonts,amsmath,amsthm}
\usepackage{mathrsfs}
\usepackage[hidelinks]{hyperref}
\usepackage{graphicx}
\usepackage{caption}
\usepackage{subcaption}
\usepackage{enumitem}

\newcommand{\vect}[1]{\mathbf{#1}}                    
\newcommand{\vects}[1]{\pmb{#1}}                      
\newcommand{\vectNF}[1]{\hat{\mathbf{#1}}}            
\newcommand{\mat}[1]{\mathbb{#1}}                     
\newcommand{\varNF}[1]{\tilde{#1}}                    

\providecommand{\abs}[1]{\lvert#1\rvert}              
\newcommand{\opt}[1]{\pmb{\mathcal{#1}}^{\dag}}       
\newcommand{\opts}[1]{\mathcal{#1}^{\dag}}            
\newcommand{\normp}[1]{\left(#1\right)}               
\newcommand{\normb}[1]{\left[#1\right]}               

\newcommand{\onehalf}{\frac{1}{2}}
\newcommand{\dudt}[1]{\frac{\partial #1}{\partial t}} 
\newcommand{\dudx}[1]{\frac{\partial #1}{\partial x}} 

\newcommand{\refEq}[1]{Eq. (\ref{#1})}
\newcommand{\refFig}[1]{Figure \ref{#1}}
\newcommand{\refTab}[1]{Table \ref{#1}}

\begin{document}
\begin{frontmatter}
\title{Application of discrete adjoint method to sensitivity and uncertainty analysis in steady-state two-phase flow simulations}
\author{Guojun Hu\corref{corAuthor}}
\ead{ghu3@illinois.edu}
\author{Tomasz Kozlowski\corref{}}
\ead{txk@illinois.edu}
\address{Department of Nuclear, Plasma, and Radiological Engineering, University of Illinois at Urbana-Champaign\\
	Talbot Laboratory 104 S Wright St, Urbana IL, 61801, United States}

\cortext[corAuthor]{Corresponding author}

\begin{abstract}
	Verification, validation and uncertainty quantification (VVUQ) have become a common practice in thermal-hydraulics analysis. An important step in the uncertainty analysis is the sensitivity analysis of various uncertain input parameters. The common approach for computing the sensitivities, e.g. variance-based and regression-based methods, requires solving the governing equation multiple times, which is expensive in terms of computational cost. An alternative approach to compute the sensitivities is the adjoint method. The cost of solving an adjoint equation is comparable to the cost of solving the governing equation. Once the adjoint solution is available, the sensitivities to any number of parameters can be obtained with little cost. However, successful application of adjoint sensitivity analysis to two-phase flow simulations is rare. In this work, an adjoint sensitivity analysis framework is developed based on the discrete adjoint method and a new implicit forward solver. The framework is tested with the faucet flow problem and the BFBT benchmark. Adjoint sensitivities are shown to match analytical sensitivities very well in the faucet flow problem. The adjoint method is used to propagate uncertainty in input parameters to the void fraction in the BFBT benchmark test. The uncertainty propagation with the adjoint method is verified with the Monte Carlo method and is shown to be efficient. 
\end{abstract}


\end{frontmatter}

\section{Introduction}
In recent years, verification, validation and uncertainty quantification (VVUQ) have become a common practice in thermal-hydraulics analysis. In general, these activities deal with propagation of uncertainties in computer code simulations, e.g., through system analysis codes. An important step in uncertainty analysis is the Sensitivity Analysis (SA) of various uncertain input parameters. A common approach to calculate sensitivity includes variance-based and regression-based methods. However, these methods require solving the system of interest (in our case, two-phase flow) multiple times, sometimes 100s of times, using different input parameters, which is very expensive in terms of CPU time. An alternative approach for calculating sensitivities is the adjoint method. The cost of solving an adjoint equation is comparable to the cost of solving the governing equation (forward equation, e.g. the two-phase two-fluid model equation). However, once the adjoint solution is available, the sensitivity to an arbitrary number of parameters can be calculated at the same time.

There is a long history of using the Adjoint Sensitivity Analysis (AdSA) in optimal control theory. The use of adjoint method for computing sensitivities came up in the nuclear industry in the 1940s. Later, the adjoint method became popular in computational fluid dynamics field \cite{Marchuk1995, Giles2000}. Within the field of aeronautical computational fluid dynamics, the use of adjoint method has been seen in \cite{Jameson1988, Jameson1994, Jameson1998, Nadarajah2000}. Adjoint problems arise naturally in the formulation of methods for optimal aerodynamic design and optimal error control \cite{Giles1998,Giles2000, Giles2001, Giles2003}. Adjoint solution provides the linear sensitivities of an objective function (e.g. lift force and drag force) to a number of design variables. These sensitivities can then be used to drive an optimization procedure. In a sequence of papers, Jameson developed the adjoint approach for the potential flow, the Euler equation, and the Navier-Stokes equation \cite{Jameson1988, Jameson1994, Jameson1998, Nadarajah2000}. Many of these methods were based on the continuous form of the governing equation. These methods belong to the group of so-called continuous adjoint method \cite{Marchuk1995}.

The application of the adjoint method to optimal aerodynamic design is very successful. However, to the author's best knowledge, successful application of AdSA to two-phase flow problems is rare. Cacuci performed an AdSA to two-phase flow problems using the RELAP5/MOD3.2 numerical discretization \cite{Cacuci1982, Cacuci2000a, Cacuci2000b}. This method belongs to a group of so-called discrete adjoint method \cite{Marchuk1995}. An application of Cacuci's approach was illustrated by \cite{petruzzi2008development}, where the approach was applied to the blow-down of a gas from a pressurized vessel. In previous work, the author applied a continuous AdSA to steady-state two-phase flow simulations \cite{Hu2018PhdThesis}. The continuous AdSA is based on the continuous form of the governing equation. The continuous AdSA was shown to work well; however, it was based on an explicit forward solver, it was complicated in terms of derivations, and it had limitations in obtaining local sensitivity information. The objective of this paper is to develop an AdSA framework using the discrete adjoint method, which uses the discretized form of the governing equation. At first, an implicit forward solver is built based on an approximate Riemann solver. Then, an AdSA framework is developed based on the forward solver. Finally, numerical tests with faucet flow problem and the BFBT benchmark are performed to verify the framework. Adjoint sensitivities are shown to match analytical sensitivities very well in the faucet flow problem. The adjoint method is then used to propagate uncertainty in input parameters to the void fraction in the BFBT benchmark test. The uncertainty propagation with the adjoint method is verified with the Monte Carlo method. 

This article is organized in the following way. Section \ref{sec-two} presents briefly the forward numerical scheme for solving the two-phase two-fluid model equation. Section \ref{sec-three} presents the discrete AdSA framework. Section \ref{sec-four} presents the numerical tests for verification and assessment of discrete AdSA. Section \ref{sec-five} concludes the current work.

\section{Forward solver}\label{sec-two}
\subsection{Two-phase two-fluid model}\label{sec-two-p1}
For 1D problems, the basic two-phase two-fluid six-equation model without any differential closure correlations \cite{ishii2010thermo} can be written in a vector form as
\begin{equation}\label{Eq-A.05}
    \dudt{\vect{U}} + \dudx{\vect{F}} + \vect{P}_{ix}\dudx{\alpha_g} + \vect{P}_{it}\dudt{\alpha_g} = \vect{S}
\end{equation}
where $\vect{U}$ is the vector of conservative variables, $\vect{F}$ is the vector of flux variables, $\vect{P}_{ix}$ and $\vect{P}_{it}$ are the vectors related to the partial derivatives of the void fraction, and $\vect{S}$ is the vector of source terms. They are defined as
 \begin{align}\label{Eq-A.06}
    \vect{U}& \equiv \begin{pmatrix}
                \alpha_l\rho_l \\
                \alpha_l\rho_l u_l \\
                \alpha_l\rho_l E_l\\
                \alpha_g\rho_g \\
                \alpha_g\rho_g u_g \\
                \alpha_g\rho_g E_g\\
                \end{pmatrix},
     \vect{F} \equiv \begin{pmatrix}
                \alpha_l\rho_l u_l \\
                \alpha_l\rho_l u_l^2 + \alpha_l p \\
                \alpha_l\rho_l H_l u_l\\
                \alpha_g\rho_g u_g\\
                \alpha_g\rho_g u_g^2 + \alpha_g p \\
                \alpha_g\rho_g H_g u_g\\
                \end{pmatrix}\\
      \vect{W} &\equiv \begin{pmatrix}
                \alpha_g \\
                p \\
                T_l\\
                T_g \\
                u_l \\
                u_g\\
                \end{pmatrix}, \vect{P}_{ix} \equiv \begin{pmatrix}
                        0 \\
                        p \\
                        0 \\
                        0 \\
                       -p \\
                        0\\
                    \end{pmatrix},
     \vect{P}_{it} \equiv \begin{pmatrix}
                         0 \\
                         0 \\
                        -p \\
                         0 \\
                         0 \\
                         p\\
                    \end{pmatrix}
\end{align}
Let the subscript $k=l,g$ denote the liquid phase and gas phase, respectively. The variables $\normp{\alpha_k, \rho_k, u_k, e_k}$ denote the volume fraction, the density, the velocity, and the specific internal energy of $k$-phase. The summation of phasic volume fraction should be one, i.e. $\alpha_l + \alpha_g = 1$.  $p$ is the pressure of two phases. $E_k = e_k + u_k^2/2$ and $H_k = e_k + p/\rho_k + u_k^2/2$ are the phasic specific total energy and specific total enthalpy.

An appropriate Equation of State (EOS) is required to close the system. For many practical problems in the nuclear thermal-hydraulics analysis, the temperature of two phases are required to model the source terms. In such a case, a useful EOS is given by specifying the Gibbs free energy  as a function of pressure and temperature $T_k$, i.e.
\begin{equation}
\mathfrak{g}_k = \mathfrak{g}_k(T_k,p), \text{  for  }k=l,g
\end{equation}
After specifying the specific Gibbs free energy, the phasic density and specific internal energy are obtained from the partial derivatives of the specific Gibbs free energy. The details about specifying the EOS through the specific Gibbs free energy are referred to \cite{iapws1998, Hu2018PhdThesis}.

Closure correlations are required for simulating the system behavior of a boiling pipe. Closure correlations based on RELAP5-3D code manual \cite{RELAP5V4} are used to model the source vector. The details are referred to \cite{Hu2017Riemann, Hu2018PhdThesis}.
\subsection{Numerical scheme}\label{sec-two-p2}
\begin{figure}[!htb]
  \centering
  \includegraphics[width=0.45\textwidth]{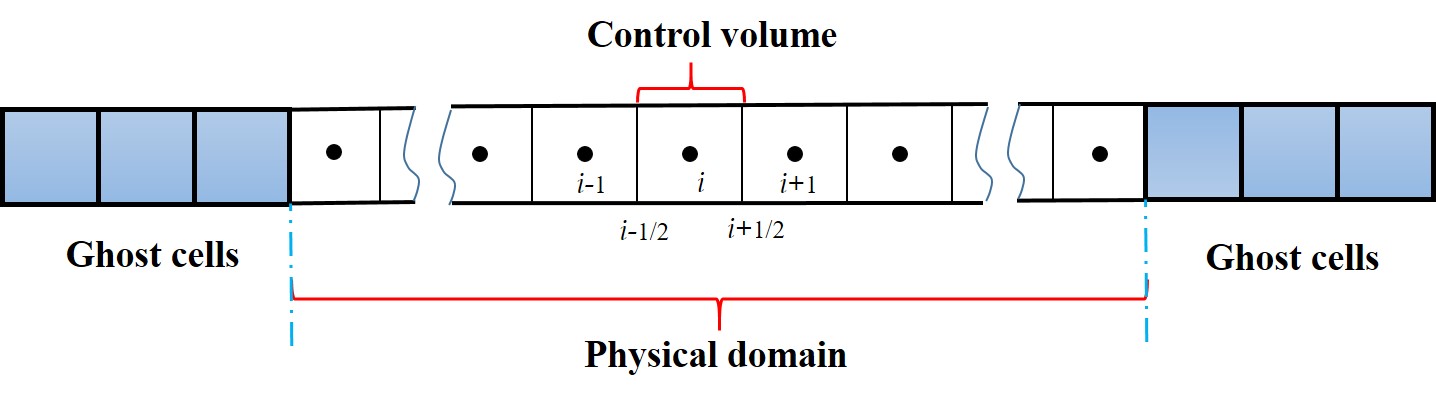}\\
  \caption{Schematic of the 1D spatial discretization}\label{forward:refFig-A1}
\end{figure}

For 1D problems, the spatial discretization is shown schematically in \refFig{forward:refFig-A1}. The physical domain is divided into $N$ cells. The cell center is denoted with an index $i$ and the cell boundaries are denoted with $i\pm 1/2$, for $i=1,\cdots, N$. All unknown variables are solved in the cell center (collocated mesh). On each side of the physical domain, ghosts cells are used to deal with boundary conditions.

The two-phase two-fluid model is solved with the backward Euler method, which gives a fully implicit scheme
\begin{equation}\label{backward-euler-fully-implicit}
    \vect{U}_i^{n+1}-\vect{U}_i^{n} + \vect{P}_{it,i}^{n+1}\normp{\alpha_{g,i}^{n+1}-\alpha_{g,i}^{n}}=  \Delta t \opt{L}\normp{\vect{U}_i^{n+1}}
\end{equation}
where $\opt{L}$ is an operator representing the spatial differential operators and the source terms,
\begin{equation}\label{backward-euler-operator}
      \opt{L}\normp{\vect{U}_i}= - \frac{\vectNF{F}_{i+1/2} -\vectNF{F}_{i-1/2}}{\Delta x}-\vect{P}_{ix,i}\frac{\alpha_{g,i+1}-\alpha_{g,i-1}}{2\Delta x} + \vect{S}_i
\end{equation}
where $\vectNF{F}_{i+1/2}$ and $\vectNF{F}_{i-1/2}$ are low-order numerical fluxes at cell boundaries. \refEq{backward-euler-fully-implicit} forms a set of algebraic nonlinear equations, which is solved with the JFNK method. More details of the numerical scheme can be seen in \cite{Hu2018Implicit}.

A first-order Roe-type numerical flux is constructed following the Roe-Pike \cite{toro2013, glaister1988approximate} method. Let $\mat{A}_c$ be the Jacobian matrix defined as $\mat{A}_c \equiv \partial \vect{F}/\partial \vect{U}$. The subscript $c$ denotes that the Jacobian matrix and eigenvalues/eigenvectors are obtained with the conservative part of the equation \cite{Hu2017Riemann, Hu2018PhdThesis}. Let $\lambda_{c}^{m}$ and $\vect{K}_{c}^{m}$, for $m = 1, \cdots, 6$ be the eigenvalues and right eigenvectors of the matrix $\mat{A}_c$. Let $Q(z)$ be a scalar function defined as
\begin{equation}
    Q(z) = \left\{
     \begin{array}{l}
       \frac{1}{2}\normp{\frac{z^2}{\delta} + \delta}, \quad \abs{z} < \delta \\
       \abs{z}, \quad \quad \quad \quad \abs{z} \geq \delta
     \end{array}\right.
\end{equation}
where $\delta$ is the coefficient for the addition of numerical viscosity term, which is set at 0.125 as was used by Yee (\cite{Yee1985}).  The Roe-type numerical flux is constructed by
\begin{equation}\label{R-L-Flux}
    \vectNF{F}_{i+1/2} = \onehalf\normp{\vect{F}_i +\vect{F}_{i+1}} - \onehalf \sum_{m=1}^{6}\varNF{\omega}_{i+1/2}^{m} Q\normp{\varNF{\lambda}_{c, i+1/2}^{ m}}\varNF{\vect{K}}_{c, i+1/2}^{m}
\end{equation}
where $\omega_{i+1/2}^{m}$ is the wave strength when projecting the conservative vectors to the characteristic space. The Jacobian matrix, approximate eigenvalues, and eigenvectors are given in the Appendix A for reference. Complete details about the approximate eigenvalues/eigenvectors and the average state are referred to \cite{Hu2017Riemann, Hu2018PhdThesis}. 

\subsection{Jacobian-Free Newton-Krylov method}\label{sec-three-p3}
\refEq{backward-euler-fully-implicit} forms a set of algebraic nonlinear equations, which are solved  with the JFNK method. \refEq{backward-euler-fully-implicit} can be generalized as
\begin{equation}
    \vect{G}(\vect{W}) = \vect{0}
\end{equation}
where $\vect{R}$ denotes the global residual function of \refEq{backward-euler-fully-implicit} and $\vect{W}$ is the vector of unknown primitive variables. The JFNK method is based on the Newton's method, which solves the nonlinear algebraic equations iteratively
\begin{align}
    \mat{J}^{m}\delta \vect{W}^{m} &= - \vect{G}(\vect{W}^{m}) \label{JFNK-linear-eq} \\
    \vect{W}^{m+1} &= \vect{W}^{m} + \delta \vect{W}^{m}\label{JFNK-update}
\end{align}
where $m$ denotes the $m$-th step of the iteration. The iteration starts with an initial guess of $\vect{W}$ which is usually taken from the old time step. $\mat{J}^{m}$ is the Jacobian matrix define as
\begin{equation}
    \mat{J}^{m} \equiv \normp{\frac{\partial \vect{G}}{\partial \vect{W}}}^{m}
\end{equation}
In the JFNK scheme, the linear equation \refEq{JFNK-linear-eq} is solved with the Krylov subspace method. The essential idea of the JFNK method is that the Krylov method requires only the matrix-vector product and the explicit form of the Jacobian matrix could be avoided. The matrix-vector project is approximated with
\begin{equation}
    \mat{J}^{m}\vect{v} \approx \frac{\vect{G}(\vect{W}^{m} + \epsilon \vect{v})-\vect{G}(\vect{W}^{m})}{\epsilon}
\end{equation}
where $\vect{v}$ is the Krylov vector and $\epsilon$ is a small parameter. In this article, the JFNK method is implemented with the scientific computational toolkit PETSc \cite{balay2016petsc}.

\section{Adjoint sensitivity analysis}\label{sec-three}
\subsection{General framework}\label{sec-three-p1}
Let $\opt{G}$ be the operator that represents the governing equation of the forward problem, e.g. the two-phase two-fluid equation. Let $\vect{W}$ be the vector of physical variables. For the forward problem, there are usually a few input parameters, denoted by $\vects{\omega}$, that affect the flow field, e.g. boundary conditions and physical model parameters. Suppose the governing equation is written as
\begin{equation}\label{forward-governing-eq}
    \opt{G}\normp{\vect{W}, \vects{\omega}} = \vect{0}
\end{equation}
Let $\opts{R}$ be the operator that measures the interested quantity, objective function, or response $R$, e.g. void fraction.  The response could be expressed as
\begin{equation}\label{forward-response}
    R = \opts{R}\normp{\vect{W}, \vects{\omega}}
\end{equation}

In the following discussion, vectors and matrices are defined in a way such that the multiplications shown in the following equations are the inner product. \refEq{forward-governing-eq} gives
\begin{equation}\label{adjoint:refEq-A1}
     \frac{\partial \opt{G}}{\partial \vect{W}}\frac{\mathrm{d}\vect{W}}{\mathrm{d}\vects{\omega} } + \frac{\partial \opt{G}}{\partial \vects{\omega}}= \vect{0}
\end{equation}
\refEq{forward-response} gives
\begin{equation}\label{adjoint:refEq-A2}
    \frac{\mathrm{d}R}{\mathrm{d}\vects{\omega}} = \frac{\partial \opts{R}}{\partial \vect{W}}\frac{\mathrm{d}\vect{W}}{\mathrm{d}\vects{\omega} } + \frac{\partial \opts{R}}{\partial \vects{\omega}}
\end{equation}
where $\mathrm{d}R/\mathrm{d}\vects{\omega}$ is the sensitivity of interest. In the following discussions, the partial derivatives in \refEq{adjoint:refEq-A1} and \refEq{adjoint:refEq-A2} are called coefficient matrices/vectors. \refEq{adjoint:refEq-A2} depends on $\mathrm{d}\vect{W}/\mathrm{d}\vects{\omega}$, which is usually expensive to obtain as it involves solving the forward governing equations. The idea in the adjoint method is to remove the dependency on $\mathrm{d}\vect{W}/\mathrm{d}\vects{\omega}$ by combining \refEq{adjoint:refEq-A1} and \refEq{adjoint:refEq-A2} using the Lagrange multiplier approach.

Let $\vects{\phi}$ be the vector of Lagrange multiplier, which is a vector of free variables. Multiplying \refEq{adjoint:refEq-A1} with the transpose of $\vects{\phi}$ and subtracting the result from \refEq{adjoint:refEq-A2}, the sensitivity of interest is transformed into
\begin{equation}
    \frac{\mathrm{d}R}{\mathrm{d}\vects{\omega}} = \normp{\frac{\partial \opts{R}}{\partial \vect{W}}-\vects{\phi}^{T}\frac{\partial \opt{G}}{\partial \vect{W}}} \frac{\mathrm{d}\vect{W}}{\mathrm{d}\vects{\omega} }+ \normp{\frac{\partial \opts{R}}{\partial \vects{\omega}}- \vects{\phi}^{T}\frac{\partial \opt{G}}{\partial \vects{\omega}}}
\end{equation}
where the superscript $T$ denotes the transpose operator. Because the Lagrange multiplier $\vects{\phi}$ is a vector of free variables, it can be chosen in a way such that
\begin{equation}\label{adjoint:refEq-A5}
    \frac{\partial \opts{R}}{\partial \vect{W}}-\vects{\phi}^{T}\frac{\partial \opt{G}}{\partial \vect{W}} = \vect{0}
\end{equation}
The so-called adjoint equation is obtained by taking the transpose of \refEq{adjoint:refEq-A5}, i.e.
\begin{equation}\label{general-adjoint-equation}
    \normp{\frac{\partial \opt{G}}{\partial \vect{W}}}^{T}\vects{\phi}= \normp{\frac{\partial \opts{R}}{\partial \vect{W}}}^{T}
\end{equation}
The Lagrange multiplier $\vects{\phi}$ given by \refEq{general-adjoint-equation} is the adjoint solution. After obtaining the adjoint solution, the sensitivity of interest is obtained with
\begin{equation}\label{general-adjoint-sensitivity}
    \frac{\mathrm{d}R}{\mathrm{d}\vects{\omega}} =  \frac{\partial \opts{R}}{\partial \vects{\omega}}- \vects{\phi}^{T}\frac{\partial \opt{G}}{\partial \vects{\omega}}
\end{equation} 
Note that if the response function does not depend explicitly on $\vects{\omega}$, then $\partial\opts{R}/\partial\vects{\omega}$ can be removed from \refEq{general-adjoint-sensitivity}. The advantage of \refEq{general-adjoint-sensitivity} is that it is independent of $\delta\vect{W}$, which means that the sensitivity of the response to an arbitrary number of parameters can be determined without the need for additional forward calculations.

\subsection{Discrete adjoint sensitivity analysis to two-fluid model}
The adjoint equation \refEq{general-adjoint-equation} is problem dependent. In this paper, the general AdSA framework is applied to the two-phase two-fluid model for steady-state problems. At steady-state, the governing equation reduces to
\begin{equation}\label{adjoint:refEq-A8}
    \dudx{\vect{F}} + \vect{P}_{ix}\dudx{\alpha_g} -\vect{S}=\vect{0}
\end{equation}
The discretized form of \refEq{adjoint:refEq-A8} is thus
\begin{equation}\label{forward-governing-eq-ss-discrete}
    \vect{G}_i\normp{\vect{W}, \vects{\omega}} = -\frac{\vectNF{F}_{i+1/2} -\vectNF{F}_{i-1/2}}{\Delta x}-\vect{P}_{ix,i}\frac{\alpha_{g,i+1}-\alpha_{g,i-1}}{2\Delta x} + \vect{S}_i
\end{equation}
Let $\vect{G}(\vect{W}, \vects{\omega})$ be the global residual vector that is assembled from \refEq{forward-governing-eq-ss-discrete}. The operator $\opt{G}$ in \refEq{forward-governing-eq} is then defined as
\begin{equation}
    \opt{G}\normp{\vect{W}, \vects{\omega}} \equiv \vect{G}(\vect{W}, \vects{\omega})
\end{equation}
Let $R(\vect{W})$ be the scalar function that represents the response function of interest

The response operator $\opts{R}$ is then defined as
\begin{equation}
    \opts{R} \equiv R\normp{\vect{W}}
\end{equation}
Since the operator $\opt{G}$ and $\opts{R}$ now represent a global residual vector and a scalar function, the coefficient matrices/vectors in \refEq{general-adjoint-equation} and \refEq{general-adjoint-sensitivity} are well defined. \refEq{general-adjoint-equation} and \refEq{general-adjoint-sensitivity} are thus transformed into
\begin{equation}\label{discrete-adjoint-equation-ss}
    \normp{\frac{\partial \vect{G}}{\partial \vect{W}}}^{T}\vects{\phi}= \normp{\frac{\partial R}{\partial \vect{W}}}^{T}
\end{equation}

\begin{equation}\label{discrete-adjoint-sensitivity-ss}
    \frac{\mathrm{d}R}{\mathrm{d}\vects{\omega}} = - \vects{\phi}^{T}\frac{\partial \vect{G}}{\partial \vects{\omega}}
\end{equation}
In the following discussion, the sensitivities obtained with the discrete AdSA method will be denoted with `DAS'. 

Because of the nonlinearity in the governing equation, it is impractical to obtain analytical coefficient matrices/vectors required by \refEq{discrete-adjoint-equation-ss} and \refEq{discrete-adjoint-sensitivity-ss}. Numerical differentiation using a finite difference method is used to calculate the coefficient matrices/vectors, including $\partial \vect{G}/\partial \vect{W}$, $\partial R/\partial \vect{W}$, $\partial \vect{G}/\partial \vects{\omega}$, and $\partial R/\partial \vects{\omega}$. 

\subsection{Perturbation method}
Unless for simple problems, it is usually impractical to obtain analytical sensitivity for verifying the adjoint sensitivity. In this paper, the response is taken as a linear function of the unknown primitive variables (i.e. $\vect{W}$), the sensitivity can be obtained by solving for $\mathrm{d}\vect{W}/\mathrm{d}\vects{\omega}$ with \refEq{adjoint:refEq-A1}, which is the perturbation equation for the general operators $\opt{G}$ and $\opts{R}$. This is the so-called perturbation method. The discretized form of \refEq{adjoint:refEq-A1} is
\begin{equation}\label{discrete-perturbation-equation}
	\frac{\partial \vect{G}}{\partial \vect{W}}\frac{\mathrm{d}\vect{W}}{\mathrm{d}\vects{\omega}}= -\frac{\partial \vect{G}}{\partial \vects{\omega}}
\end{equation} 
In the following discussion, the sensitivities obtained with the perturbation method will be denoted with `PS'. 

The computational effort to solve the perturbation equation, \refEq{discrete-perturbation-equation}, is similar to solve the adjoint equation, \refEq{discrete-adjoint-equation-ss}. For problems where the sensitivity can not be obtained analytically, the perturbation method is used to calculate the reference sensitivity. Let $\mathrm{M}$ denotes the total number of input parameters, which can be very large. The advantage of the adjoint method compared to the perturbation method is: for each response, the adjoint method requires solving the adjoint equation 1 time to obtain all sensitivities, while the perturbation method requires solving the perturbation equation $\mathrm{M}$ times.


\subsection{Response function}
The form of the response function does not affect the application of the adjoint method and should depend on different problems. This article considers a group of response function that can be written as
\begin{equation}
R(x_d) = \sum_{i=1}^{\mathrm{N}} \xi(x_i; x_d)q_i
\end{equation}
where $\mathrm{N}$ is the total number of cells, $x_i$ is the location of the $i^{\mathrm{th}}$ cell, and $q$ denotes a function of the primitive variable, e.g. $q = \alpha_g$. The other variable, $\xi(x)$, is a weight function which is used to study the behavior of the response function at different location $x_d$. The dependency of the response function on $x_d$ is designed for verification purposes. In this study, $\xi(x)$ is non-zero only in the neighboring cells of $x_d$, which simulates a point-wise response function.

\section{Numerical tests}\label{sec-four}
\subsection{Faucet flow}\label{sec-four-p1}
\subsubsection{Problem description}
This test is the Ransom's faucet flow problem \cite{dinh2003understanding, zou2015, hewitt1986, hewitt2013}, which has an analytical solution. This test problem consists of a liquid stream entering a vertical tube at the top and falling under gravity to form a liquid stream of decreasing cross-section. The length of the vertical tube is $L = 12$ m. Initial and boundary conditions are listed in \refTab{IC-BC-faucet-flow}. Properties of liquid and gas are obtained from the IAPWS-IF97 formulation \cite{iapws1998}. Since there is no mass and heat transfer between the liquid and gas phases, the superheated steam is used to simulate the gas phase. The source vector of this problem is
\begin{equation}
\vect{S}=\begin{pmatrix}
0 & \alpha_l \rho_l g & 0  & 0 & \alpha_g \rho_g g & 0 \\
\end{pmatrix}^{T}
\end{equation}
where $g$ is the gravitational acceleration constant and the superscript $T$ is the transpose operator.

\begin{table}[!htbp]
	\centering
	\caption{Initial and boundary conditions for the faucet flow problem}\label{IC-BC-faucet-flow}
	\begin{tabular}{lcc}
		\hline
		Variables  & Initial conditions & Boundary conditions \\ \hline
		$\alpha_g$ & 0.2 & $\alpha_{g, inlet} = 0.2$  \\
		$p$  (MPa) & 0.1 & $p_{outlet} = 0.1$ \\
		$T_l$ (K)  & 300. & $T_{l, inlet} = 300.$  \\
		$T_g$ (K)  & 500. & $T_{g, inlet} = 500.$  \\
		$u_l$ (m/s)& 10. & $u_{l, inlet} = 10.$  \\
		$u_g$ (m/s)& 0.  & $u_{g, inlet} = 0.$\\
		\hline
	\end{tabular}
\end{table}
The void fraction, liquid velocity, and pressure are of particular interest to this test. The steady-state solution \cite{zou2015, zou2016new} of this problem is
\begin{align}
u_{l}(x)   &= \sqrt{u_{l, inlet}^2 + 2 g_{e}x} \\
\alpha_{g}(x) &= 1.0 - \frac{\normp{1-\alpha_{g, inlet}}u_{l, inlet}}{u_{l, ss}(x)} \\
p(x)      &= p_{outlet} - \rho_g g(L - x)
\end{align}
where $g$ is the gravitational acceleration constant and $g_{e}$ is
\begin{equation}
g_{e} = g\normp{1 - \rho_g/\rho_l}
\end{equation}
where $\rho_g = 0.435$ $\mathrm{kg/m^3}$ and $\rho_l = 996.56$ $\mathrm{kg/m^3}$.

\subsubsection{Input parameters}
For this test, 3 input parameters are considered: inlet void fraction, inlet liquid velocity, and gravitational constant, i.e.
\begin{equation}
\vects{\omega} = \begin{pmatrix}
\alpha_{g,inlet} & u_{l, inlet} & g \\
\end{pmatrix}
\end{equation} 
The inlet void fraction and inlet liquid velocity represent typical input parameters related to boundary conditions; while the gravitational constant represent typical input parameters in the source terms. Several sensitivities can be obtained analytically, i.e.
\begin{equation}\label{SS-Faucet-Analytical-Sensitivity}\begin{split}
\frac{\mathrm{d}\alpha_{g}(x)}{\mathrm{d}\alpha_{g, inlet}} &= \frac{u_{l, inlet}}{u_{l}(x)}, \frac{\mathrm{d}\alpha_{g}(x)}{\mathrm{d}u_{l, inlet}} = -\frac{2\normp{1-\alpha_{g, inlet}}g_e x}{u_{l}^3(x)}, \frac{\mathrm{d}\alpha_{g}(x)}{\mathrm{d}g} = \frac{\normp{1-\alpha_{g, inlet}}u_{l, inlet} x}{u_{l}^3(x)} \\
\frac{\mathrm{d}u_{l}(x)}{\mathrm{d}u_{l, inlet}} &= \frac{u_{l, inlet}}{u_{l}(x)}, \frac{\mathrm{d}u_{l}(x)}{\mathrm{d}g} = \frac{\normp{1 - \rho_g/\rho_l}x}{u_{l}(x)},\frac{\mathrm{d}p(x)}{\mathrm{d}g} = - \rho_g\normp{L-x}
\end{split}\end{equation}
These analytical sensitivities are used to verify the AdSA framework. The sensitivity coefficient will be used for comparison, which is defined as
\begin{equation}
	\mathrm{SC} = \frac{\mathrm{d}R}{\mathrm{d}\omega}\frac{\omega_{0}}{R_{0}}
\end{equation}
where $\omega_{0}$ and $R_{0}$ are the nominal values of the input parameter and response, respectively. 

\subsubsection{Results}
The forward solver is at first run to reach steady-state for preparing the coefficient matrices/vectors. Numerical solutions match the analytical solution well. Assessment of the forward solver is ignored in this paper and is referred to \cite{Hu2018Implicit}.

\begin{figure}[!htbp]
	\centering
	\includegraphics[width=0.5\textwidth]{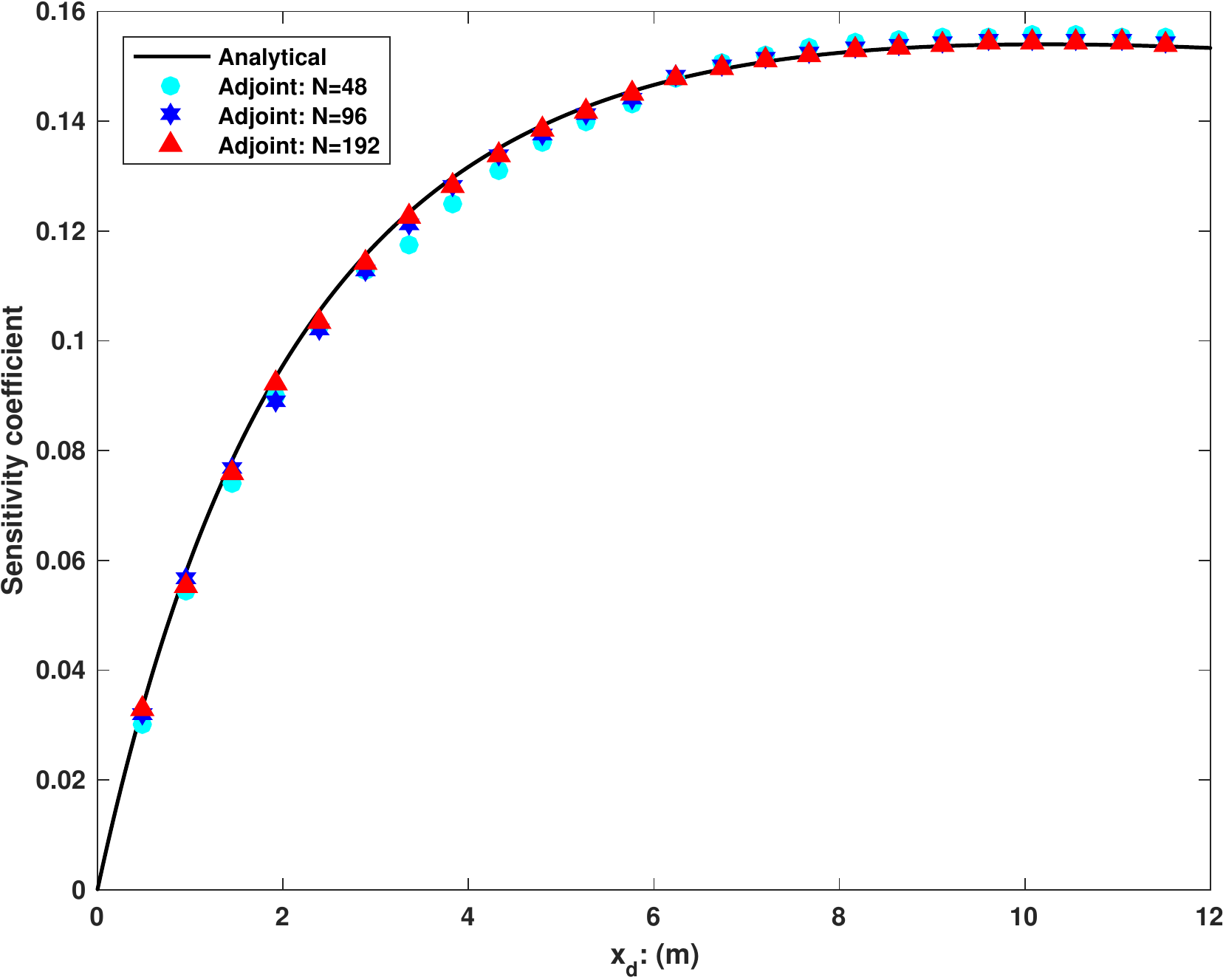}
	\caption{Effect of mesh size on the adjoint sensitivity coefficient. Sensitivity coefficient of void fraction to gravitational constant is presented.} \label{faucet-flow-sa-conv}
\end{figure}

The adjoint sensitivity is expected to depend on the discretization. A mesh convergence study is performed to decide an appropriate mesh size. Taking the void fraction and the gravitational constant as an example, \refFig{faucet-flow-sa-conv} shows the effect of mesh size on the adjoint sensitivity. As expected, the adjoint sensitivity matches the analytical one very well when the mesh is fine enough (in this case, $\mathrm{N} = 192$ is good enough). Thus, $\mathrm{N} = 192$ is used in the following analysis. 

\begin{figure}[!htbp]
	\centering
	\begin{subfigure}[t]{0.45\textwidth}
		\centering
		\includegraphics[width=\textwidth, height=0.8\textwidth]{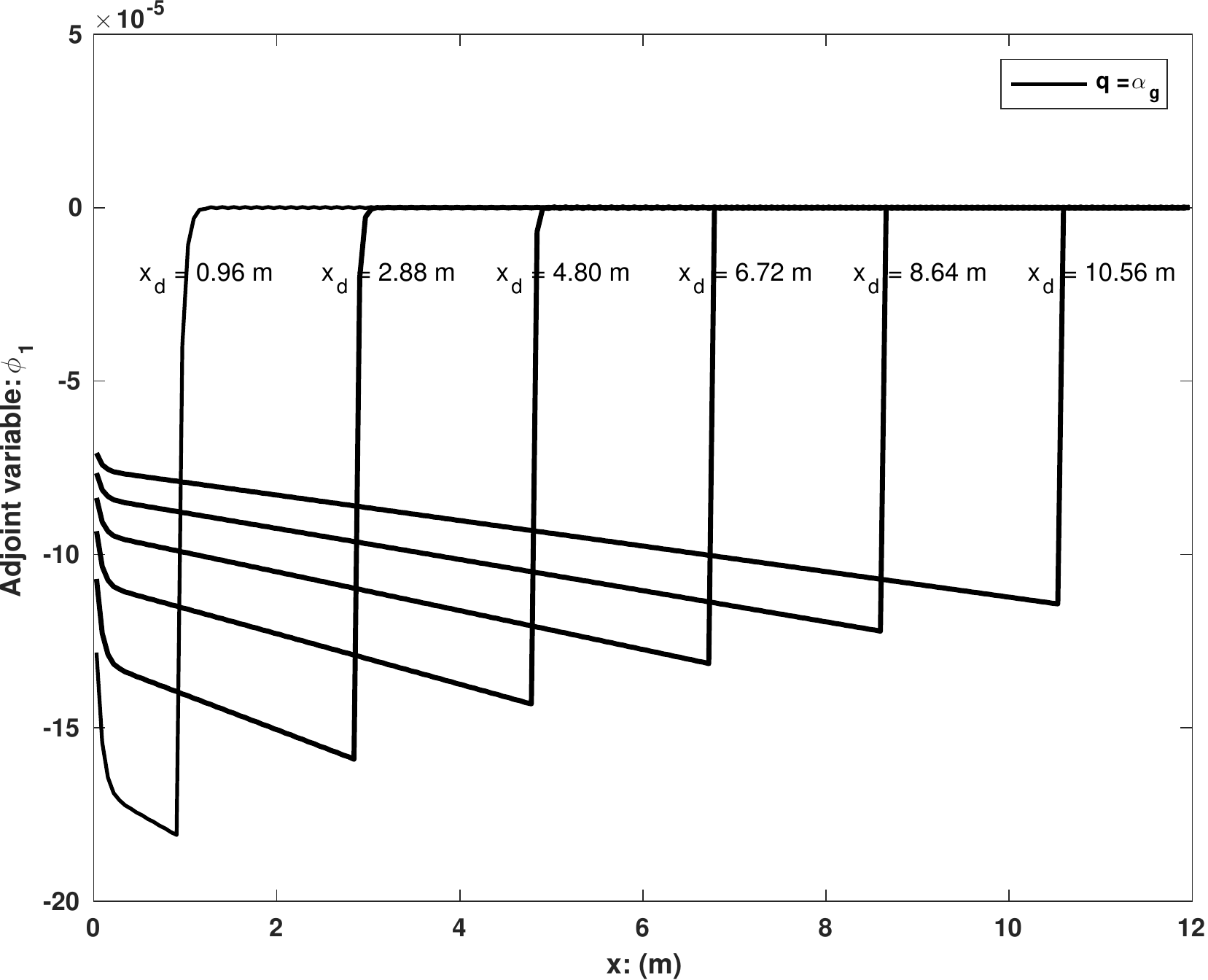}
		\caption{$\phi_1$ - Liquid mass equation for $q = \alpha_g$}
	\end{subfigure}%
	~
	\begin{subfigure}[t]{0.45\textwidth}
		\centering
		\includegraphics[width=\textwidth, height=0.8\textwidth]{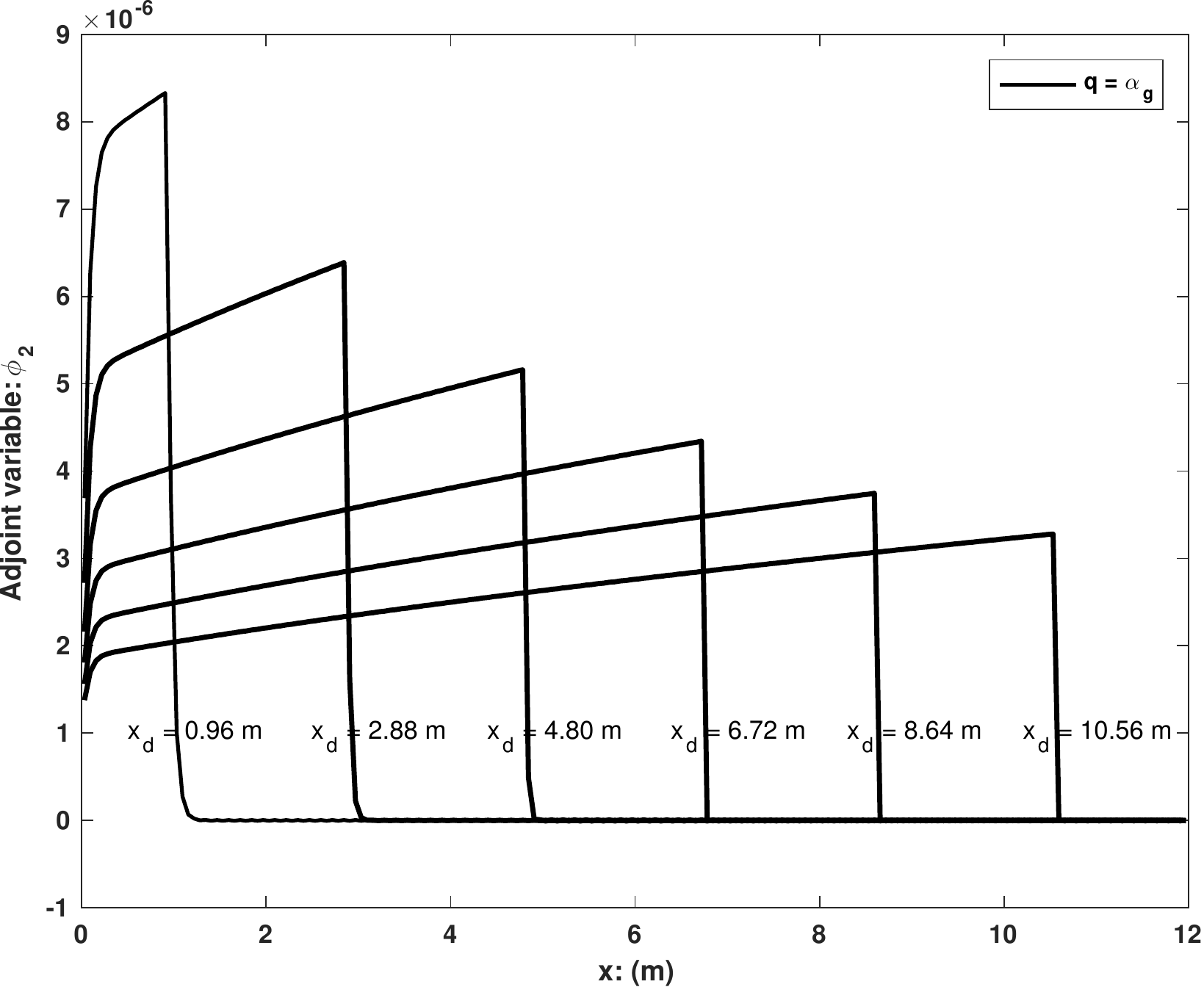}
		\caption{$\phi_2$ - Liquid momentum equation for $q = \alpha_g$}
	\end{subfigure}%

	\begin{subfigure}[t]{0.45\textwidth}
	\centering
	\includegraphics[width=\textwidth, height=0.8\textwidth]{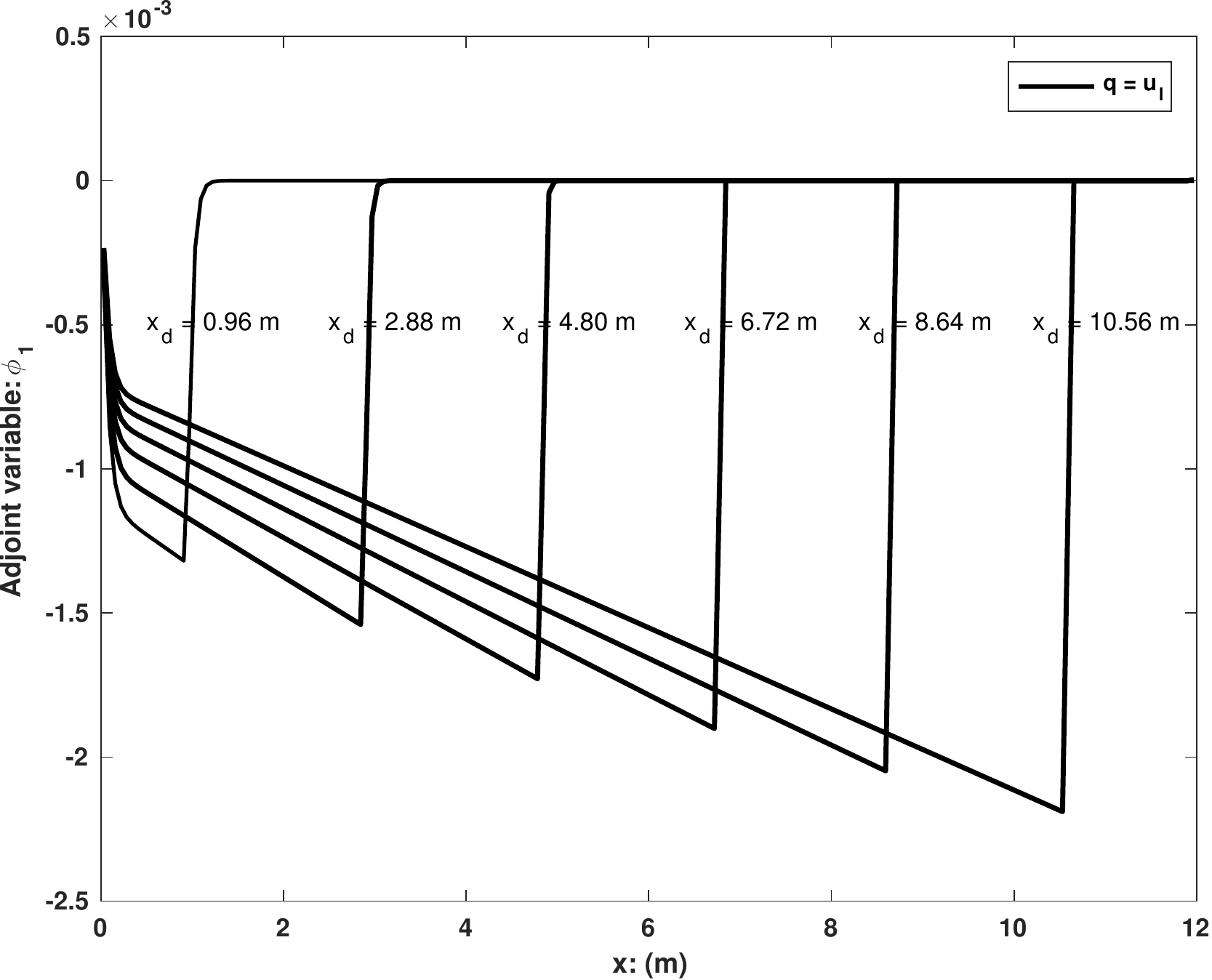}
	\caption{$\phi_1$ - Liquid mass equation for $q = u_l$}
	\end{subfigure}%
	~
	\begin{subfigure}[t]{0.45\textwidth}
	\centering
	\includegraphics[width=\textwidth, height=0.8\textwidth]{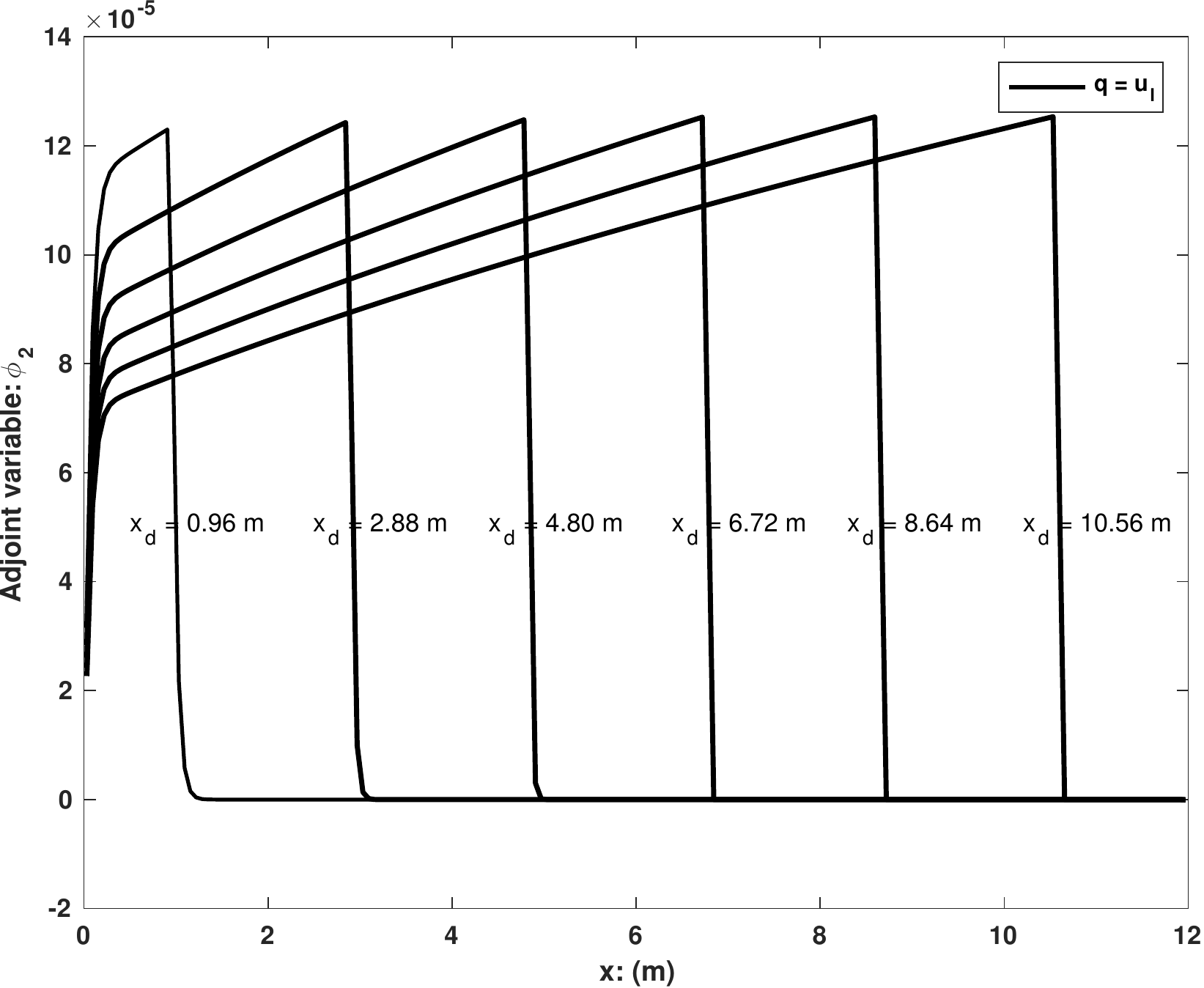}
	\caption{$\phi_2$ - Liquid momentum equation for $q = u_l$}
	\end{subfigure}%
	\caption{Adjoint solutions for responses at different locations for faucet flow problem at steady-state} \label{faucet-flow-adjoint-voidf-vl}
\end{figure}

In each cell, the adjoint solution has 6 components, representing the effect of the corresponding governing equations. For example, the first component $\phi_1$ represents the effect of the liquid mass equation. It is expected that the void fraction (and liquid velocity) are mainly affected by the liquid mass equation ($\phi_1$) and liquid momentum equation ($\phi_2$). \refFig{faucet-flow-adjoint-voidf-vl} shows examples of the adjoint components, $\phi_1$ and $\phi_2$), for responses at different locations. The profile of the adjoint components are reasonable. Because the void fraction at a particular location ($x_d$) is mainly determined by the solution at its upwind side ($x< x_d$) but not its downwind side, the adjoint components are non-zero in its upwind side but negligible in the downwind side. Note that the sharp change in the adjoint component near the inlet point is an effect of the inlet boundary conditions.

\begin{table}[!htbp]
	\centering
	\caption{Error analysis of the sensitivity coefficient from adjoint method for faucet flow at steady-state.}\label{SS-Faucet-SA-Error}
	\begin{tabular}{c|ccc|ccc|ccc}
		\hline
		& REF & $\mathrm{DAS}$ & Error: \% & REF & $\mathrm{DAS}$ & Error: \% & REF & $\mathrm{DAS}$ & Error: \%  \\ \hline
		$x_d$:(m) &  \multicolumn{3}{|c|}{$q = \alpha_g, \omega = \alpha_{g, inlet}$} &\multicolumn{3}{|c|}{$q = \alpha_g, \omega = u_{l, inlet}$}&\multicolumn{3}{|c}{$q = \alpha_g, \omega = g$} \\ \hline
		0.96	&	9.17E-01	&	9.26E-01	&	0.94	&	-1.16E-01	&	-1.10E-01	&	-5.47	&	5.82E-02	&	5.55E-02	&	-4.60	\\
		1.92	&	8.52E-01	&	8.60E-01	&	0.89	&	-1.87E-01	&	-1.83E-01	&	-1.99	&	9.33E-02	&	9.21E-02	&	-1.25	\\
		2.88	&	7.99E-01	&	8.08E-01	&	1.12	&	-2.31E-01	&	-2.26E-01	&	-1.92	&	1.15E-01	&	1.14E-01	&	-1.28	\\
		3.84	&	7.55E-01	&	7.65E-01	&	1.25	&	-2.60E-01	&	-2.55E-01	&	-1.66	&	1.30E-01	&	1.28E-01	&	-1.09	\\
		4.80	&	7.18E-01	&	7.25E-01	&	1.02	&	-2.78E-01	&	-2.76E-01	&	-1.02	&	1.39E-01	&	1.39E-01	&	-0.50	\\
		5.76	&	6.85E-01	&	6.93E-01	&	1.10	&	-2.91E-01	&	-2.88E-01	&	-0.91	&	1.45E-01	&	1.45E-01	&	-0.43	\\
		6.72	&	6.57E-01	&	6.64E-01	&	1.02	&	-2.99E-01	&	-2.97E-01	&	-0.69	&	1.49E-01	&	1.50E-01	&	0.20	\\
		7.68	&	6.32E-01	&	6.38E-01	&	1.04	&	-3.04E-01	&	-3.02E-01	&	-0.61	&	1.52E-01	&	1.52E-01	&	0.19	\\
		8.64	&	6.09E-01	&	6.16E-01	&	1.06	&	-3.06E-01	&	-3.05E-01	&	-0.52	&	1.53E-01	&	1.54E-01	&	0.20	\\
		9.60	&	5.89E-01	&	5.94E-01	&	0.86	&	-3.08E-01	&	-3.06E-01	&	-0.42	&	1.54E-01	&	1.54E-01	&	0.23	\\
		10.56	&	5.71E-01	&	5.76E-01	&	0.89	&	-3.08E-01	&	-3.07E-01	&	-0.37	&	1.54E-01	&	1.54E-01	&	0.23	\\
		11.52	&	5.54E-01	&	5.59E-01	&	0.92	&	-3.07E-01	&	-3.06E-01	&	-0.31	&	1.54E-01	&	1.54E-01	&	0.25	\\
		\hline
		$x_d$:(m) &  \multicolumn{3}{|c|}{$q = u_l, \omega = u_{l, inlet}$} &\multicolumn{3}{|c|}{$q = u_l, \omega = g$}&\multicolumn{3}{|c}{$q = p, \omega = g$} \\ \hline
		0.96	&	9.17E-01	&	9.24E-01	&	0.73	&	8.64E-02	&	8.20E-02	&	-5.04	&	-4.71E-04	&	-4.80E-04	&	1.96	\\
		1.92	&	8.52E-01	&	8.58E-01	&	0.68	&	1.61E-01	&	1.58E-01	&	-1.29	&	-4.30E-04	&	-4.37E-04	&	1.67	\\
		2.88	&	7.99E-01	&	8.07E-01	&	0.96	&	2.26E-01	&	2.22E-01	&	-1.71	&	-3.89E-04	&	-3.97E-04	&	1.97	\\
		3.84	&	7.55E-01	&	7.64E-01	&	1.14	&	2.85E-01	&	2.79E-01	&	-1.80	&	-3.48E-04	&	-3.56E-04	&	2.32	\\
		4.80	&	7.18E-01	&	7.25E-01	&	0.96	&	3.38E-01	&	3.35E-01	&	-0.81	&	-3.07E-04	&	-3.13E-04	&	1.88	\\
		5.76	&	6.85E-01	&	6.93E-01	&	1.09	&	3.87E-01	&	3.83E-01	&	-0.97	&	-2.66E-04	&	-2.72E-04	&	2.21	\\
		6.72	&	6.57E-01	&	6.63E-01	&	0.95	&	4.33E-01	&	4.32E-01	&	-0.19	&	-2.25E-04	&	-2.29E-04	&	1.49	\\
		7.68	&	6.32E-01	&	6.38E-01	&	1.03	&	4.76E-01	&	4.74E-01	&	-0.38	&	-1.84E-04	&	-1.88E-04	&	1.89	\\
		8.64	&	6.09E-01	&	6.16E-01	&	1.10	&	5.16E-01	&	5.14E-01	&	-0.51	&	-1.43E-04	&	-1.57E-04	&	9.67	\\
		9.60	&	5.89E-01	&	5.95E-01	&	0.95	&	5.55E-01	&	5.54E-01	&	-0.18	&	-1.02E-04	&	-1.03E-04	&	0.96	\\
		10.56	&	5.71E-01	&	5.76E-01	&	1.02	&	5.91E-01	&	5.89E-01	&	-0.30	&	-6.14E-05	&	-6.26E-05	&	1.84	\\
		11.52	&	5.54E-01	&	5.60E-01	&	1.08	&	6.26E-01	&	6.23E-01	&	-0.41	&	-2.05E-05	&	-2.18E-05	&	6.25	\\
		\hline
	\end{tabular}
\end{table}

\refFig{SS-Faucet-SA-Error} shows in details of the assessment of adjoint sensitivities for different combinations of the responses and input parameters. The adjoint sensitivities matches the analytical one very well, which verifies the adjoint SA framework. 

\subsection{BFBT benchmark}\label{sec-four-p2}
\subsubsection{Problem description}
One of the most valuable and publicly available databases for thermal-hydraulic modeling of Boiling Water Reactor (BWR) channels is the OECD/NEA BWR Full-size Fine-mesh Bundle Test (BFBT) benchmark, which includes sub-channel void fraction measurements in a full-scale BWR fuel assembly \cite{bfbtV1}. There are two types of void distribution measurement systems: an X-ray CT scanner and an X-ray densitometer \cite{bfbtV1}. There are 4 measurement locations, which are denoted by DEN \#3 (0.682 m), DEN \#2 (1.706 m), DEN \#1 (2.730 m), and CT (3.708 m) starting from the bottom. The geometry and system configurations of the channel are shown in \refTab{forward:TabBFBT-PC}.

\begin{table}[!htb]
	\caption{Experiment conditions for BFBT benchmark}\label{forward:TabBFBT-PC}
	\centering
	\begin{tabular}{ll|ll}
		\hline
		\multicolumn{2}{c|}{Geometry parameters} & \multicolumn{2}{c}{System/experiment parameters} \\ \hline
		Heated length (m) & 3.708 & Pressure (MPa) & 3.9 - 8.7 \\
		Width of channel box (m) & 0.1325 & Inlet temperature ($^{\circ}$C) & 238. - 292. \\
		Hydraulic diameter (m) & 0.01284  & Inlet subcooling (kJ/kg) & 50. - 56. \\
		Volumetric wall surface area ($\mathrm{m^{-1}}$) & 311.5 & Flow rate (t/h) & 10. - 70. \\
		Flow area ($\mathrm{m}^{2}$)& 9.463E-03  & Power (MW) & 0.62 - 7.3 \\
		Wetted perimeter (m) & 3.003 & Exit quality (\%) & 8. - 25. \\
		\hline
	\end{tabular}
\end{table}

Closure correlations are required for simulating the behavior of a boiling system. For this type of problems, the source vector $\vect{S}$ is modeled as
\begin{equation}
\vect{S} = \begin{pmatrix}
-\Gamma_g \\
-\alpha_l\rho_l g -f_{wl} + f_{i} - \Gamma_g u_{i} \\
Q_{wl} + Q_{il} - \Gamma_w h_{l}^{'}-\Gamma_{ig}h_{l}^{*} + \normp{f_{i}-f_{wl}-\alpha_l\rho_l g - \Gamma_g u_{i}}u_l + \Gamma_g\frac{u_l^2}{2}\\
\Gamma_g \\
-\alpha_g\rho_g g -f_{wg} - f_{i} + \Gamma_g u_{i} \\
Q_{wg} + Q_{ig} + \Gamma_w h_{g}^{'}+\Gamma_{ig}h_{g}^{*} + \normp{-f_{i}-f_{wg}-\alpha_g\rho_g g + \Gamma_g u_{i}}u_g - \Gamma_g\frac{u_g^2}{2}\\
\end{pmatrix}
\end{equation}
where $\Gamma_g$ is the net vapor generation rate due to wall vapor generation ($\Gamma_w$) and bulk vapor generation ($\Gamma_{ig}$), $u_i$ is the interface velocity, $f_i$ is the interfacial friction, $f_{wk}$ is the phasic wall friction, $Q_{ik}$ is the phasic interfacial heat flux, $Q_{wk}$ is the phasic wall heat flux, $h_k^{'}$ is the phasic enthalpy carried by the wall vapor generation term ($\Gamma_{w}$), and $h_k^{*}$ is the phasic enthalpy carried by the bulk vapor generation term ($\Gamma_{ig}$). Correlations based on RELAP5-3D code manual \cite{RELAP5V1, RELAP5V4} are used to model these variables. Details of all physical models are provided in \cite{Hu2018PhdThesis}.

\subsubsection{Input parameters}
For typical boiling problems, boundary conditions are usually specified by inlet liquid temperature, inlet liquid velocity, and outlet pressure; physical models are specified by net vapor generation rate ($\Gamma_{ig}, \Gamma_{w}$), interfacial friction ($f_i$), wall friction ($f_{wl}, f_{wg}$), interfacial heat flux ($Q_{il}, Q_{ig}$), and wall heat flux ($Q_{wl}, Q_{wg}$). Many of these physical models are correlated, e.g. $\Gamma_{ig}$, $Q_{il}$, and $ Q_{ig}$ are correlated through the interfacial heat transfer coefficients ($H_{il}, H_{ig}$). The wall heat flux ($Q_{wl}, Q_{wg}$) is closely related to the total heating power ($Q$).  Thus, there are 5 independent physical models that worth studying in details, including $f_i $,  $f_{wl}$,  $f_{wg}$, $H_{il}$, and $H_{ig}$. 

Sensitivity coefficients of the 6 primitive variables, $\alpha_g$, $p$, $T_l$, $T_g$, $u_l$, and $u_g$ to 9 parameters will be studied with the adjoint method. These 9 parameters are 
\begin{equation}\label{bfbt-input-parameters}
	\vects{\omega} = \begin{bmatrix}
             p_{outlet}  & T_{l, inlet} &  u_{l, inlet} & Q & f_i & f_{wl} & f_{wg} & H_{il} & H_{ig}  \\
             \end{bmatrix}
\end{equation}
The input parameters are perturbed by
\begin{equation}
	\omega     =  \omega_{0}\normp{1 + \varepsilon}
\end{equation}
$\omega_{0}$ and $\varepsilon$ represent the nominal value and perturbation, respectively. Perturbation in this form is mainly a numerical compromise to avoid messing with the existing closure correlations.  For the 6 physical model parameters, $\omega_{0}$  is obtained directly from the existing closure correlations and remains unchanged.

The discussion of the following tests is governed in the view of propagating uncertainties in input parameters to responses. 
 
\subsubsection{Test 1: sensitivity analysis}
\begin{figure}[!htbp]
	\centering
	\begin{subfigure}[t]{0.32\textwidth}
		\centering
		\includegraphics[width=\textwidth, height=0.8\textwidth]{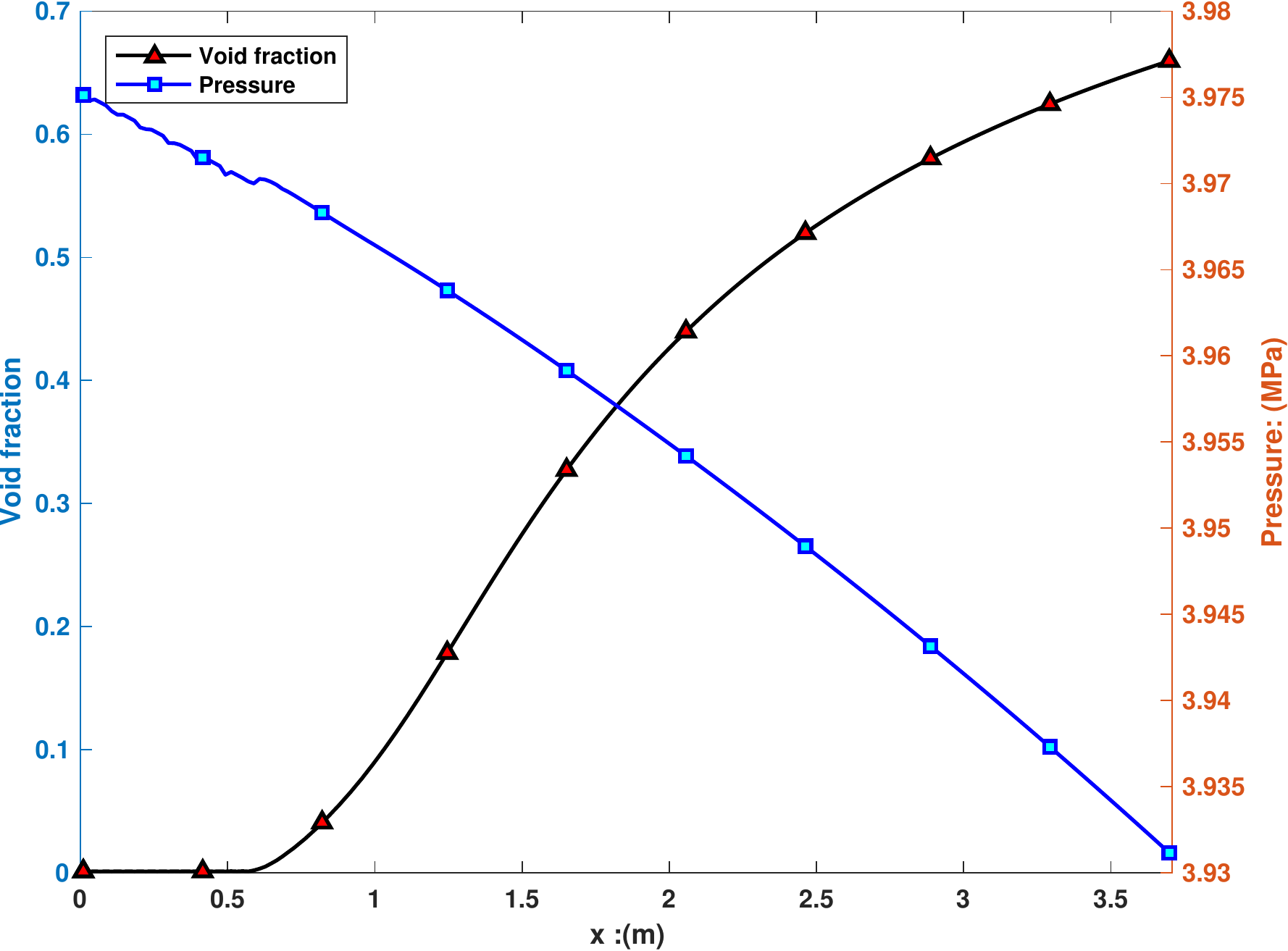}
		\caption{$\alpha_g, p$}
	\end{subfigure}%
	~
	\begin{subfigure}[t]{0.32\textwidth}
		\centering
		\includegraphics[width=\textwidth, height=0.8\textwidth]{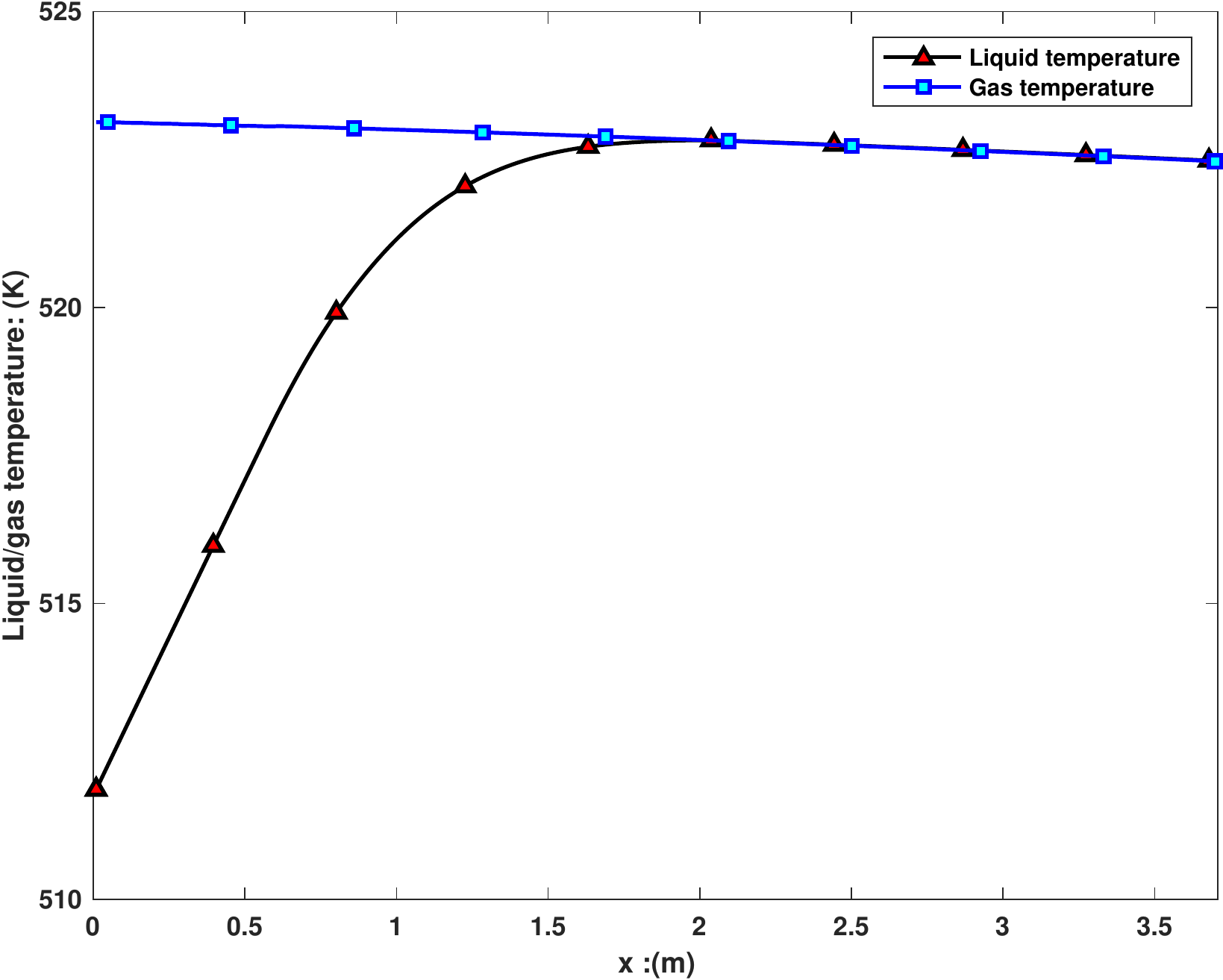}
		\caption{$T_l, T_g$}
	\end{subfigure}%
	~
	\begin{subfigure}[t]{0.32\textwidth}
		\centering
		\includegraphics[width=\textwidth, height=0.8\textwidth]{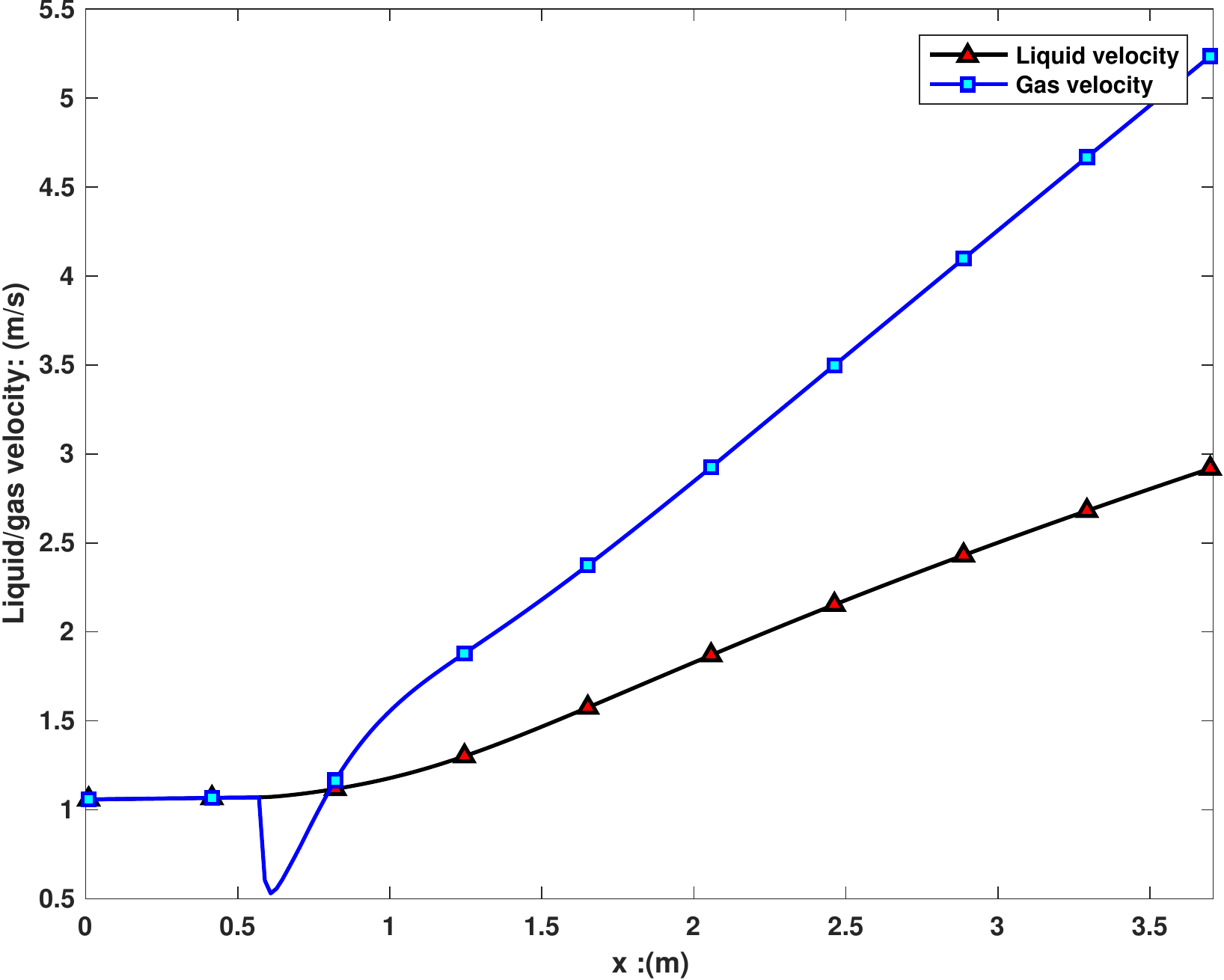}
		\caption{$u_l, u_g$}
	\end{subfigure}%
	\caption{Solution of BFBT assembly 4 case 22 at steady-state. $\mathrm{N} = 192$. } \label{SS-BFBT-A4C25-Forward-Solution}
\end{figure}

One of the test cases, assembly 4 case 22, is selected to perform a detailed adjoint SA.  The test condition for this case is: $p_{outlet} = 3.931 \text{ MPa}$, $T_{l, inlet} = 512.0 \text{ K}$, $u_{l, inlet} = 1.058 \text{ m/s}$, and assembly power = 1.57 MW. Numerically, the initial and inlet void fraction is set at $\alpha_{g, min} = 0.001$, the inlet velocities of liquid and gas phases are equal, the inlet gas temperature is equal to the saturation temperature. \refFig{SS-BFBT-A4C25-Forward-Solution} shows the steady-state forward solution of this test case, which is obtained with $\mathrm{N} = 192$. A separate test confirms that $\mathrm{N} = 192$ is fine enough for a converged solution. 

\begin{figure}[!htbp]
	\centering
	\begin{subfigure}[t]{0.32\textwidth}
		\centering
		\includegraphics[width=\textwidth, height=0.8\textwidth]{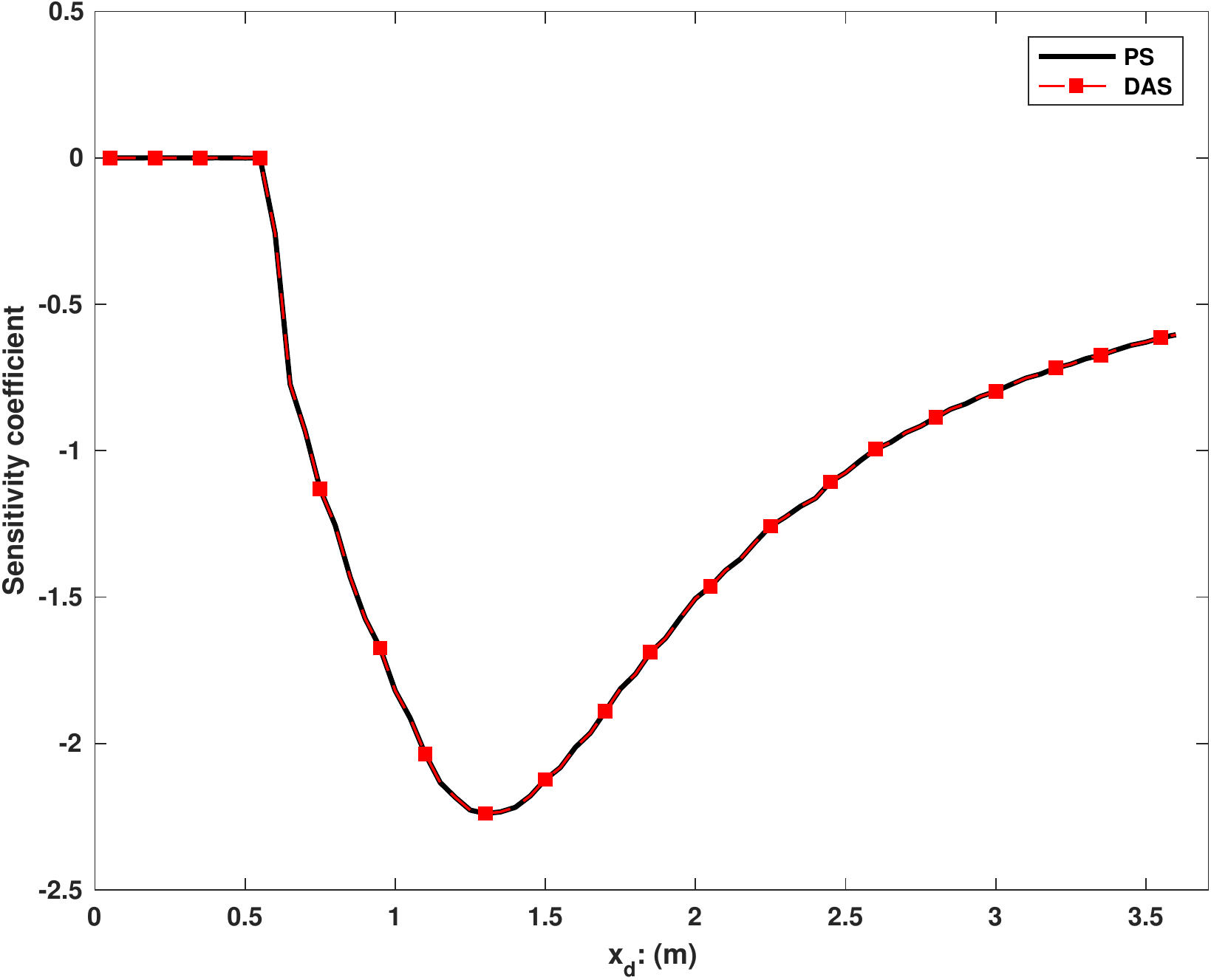}
		\caption{$q = \alpha_g, \omega = p_{outlet}$}
	\end{subfigure}%
	~
	\begin{subfigure}[t]{0.32\textwidth}
		\centering
		\includegraphics[width=\textwidth, height=0.8\textwidth]{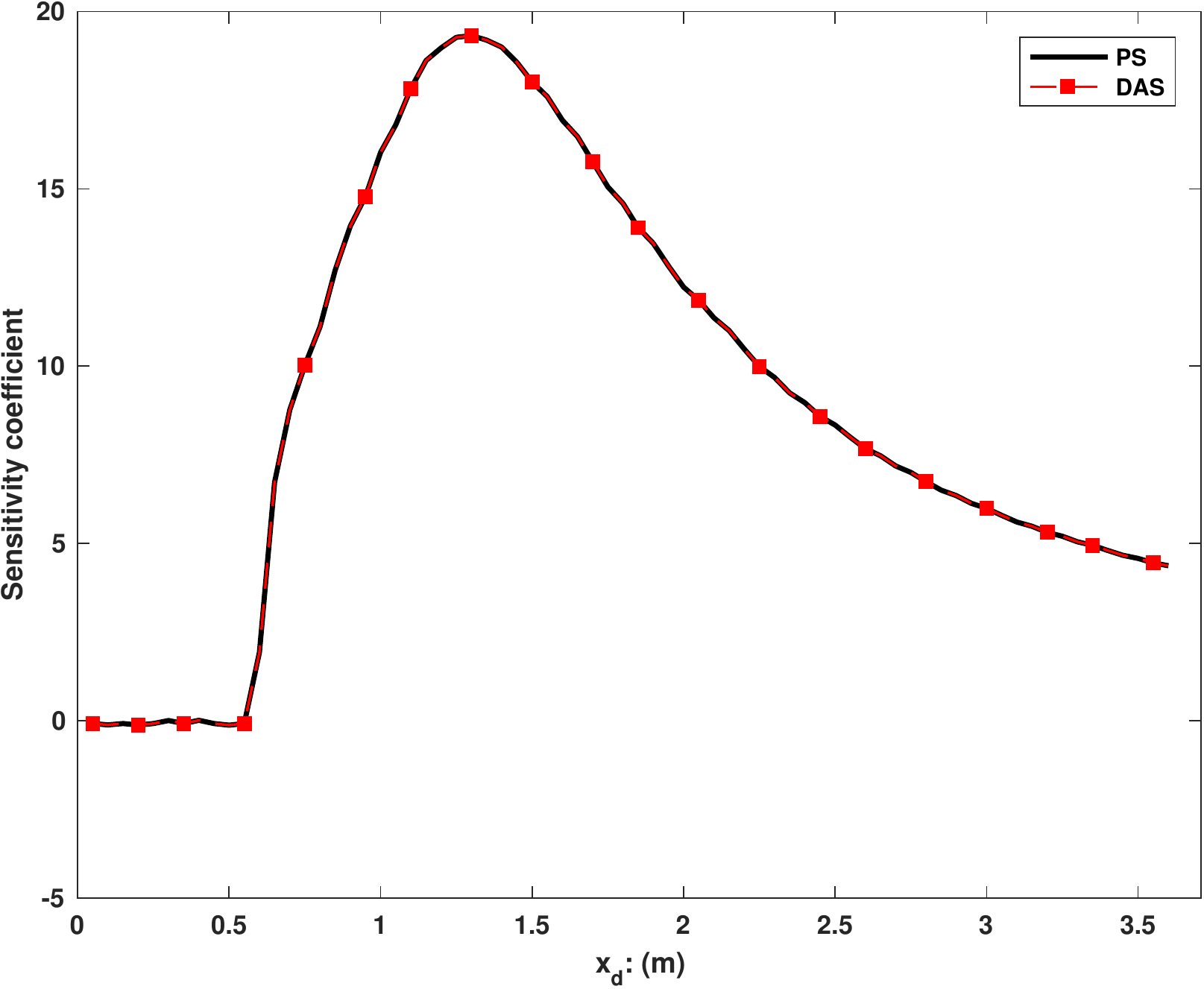}
		\caption{$q = \alpha_g, \omega = T_{l, inlet}$}
	\end{subfigure}%
	~
	\begin{subfigure}[t]{0.32\textwidth}
		\centering
		\includegraphics[width=\textwidth, height=0.8\textwidth]{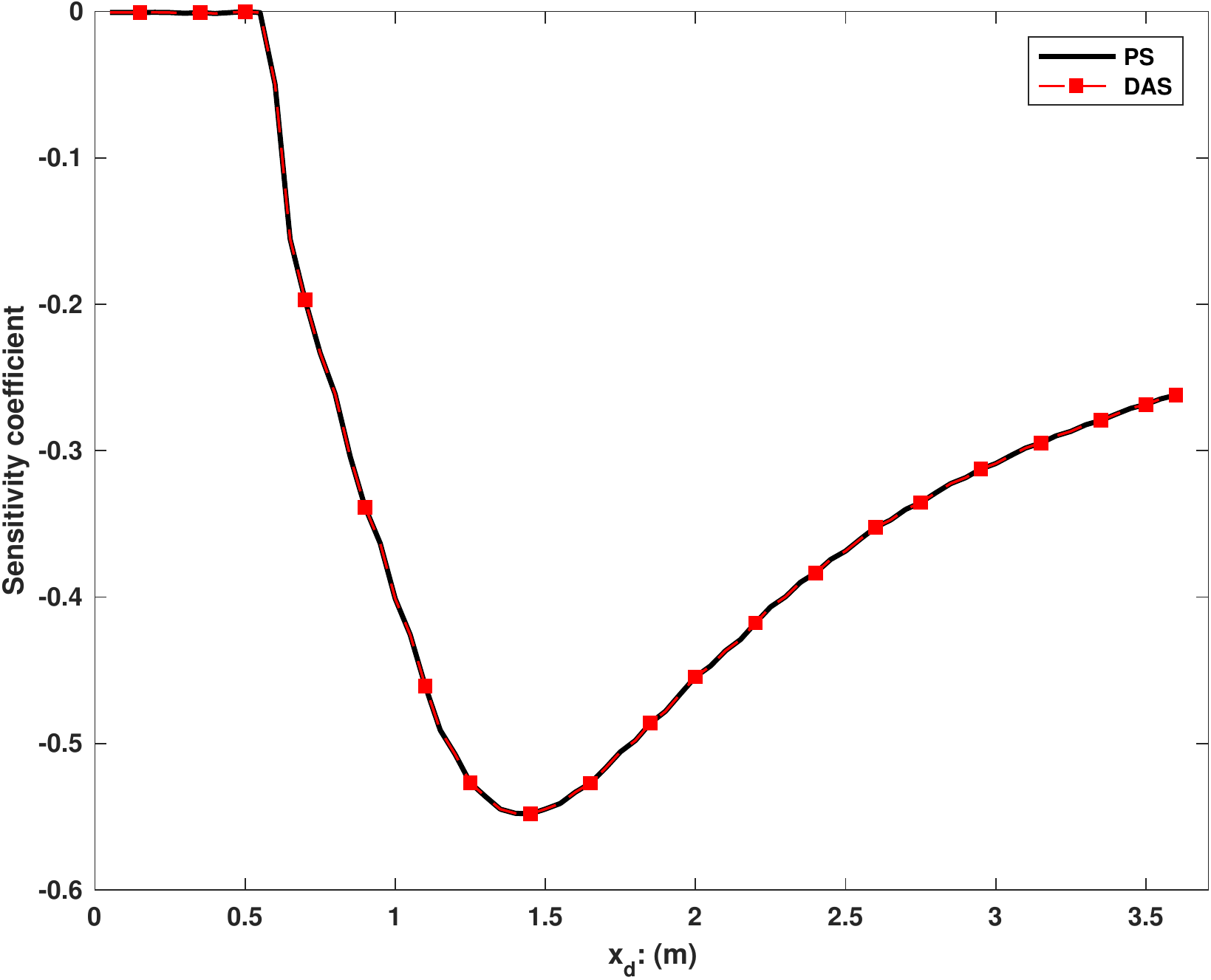}
		\caption{$q = \alpha_g, \omega = u_{l, inlet}$}
	\end{subfigure}%
	\caption{Comparison of sensitivity coefficients from discrete adjoint method (DAS) and perturbation method (PS) for BFBT assembly 4 case 22 at steady-state. $\mathrm{N} = 192$. } \label{SS-BFBT-A4C25-Sensitivity-a}
\end{figure}

\begin{figure}[!htbp]
	\centering	
	\begin{subfigure}[t]{0.32\textwidth}
		\centering
		\includegraphics[width=\textwidth, height=0.8\textwidth]{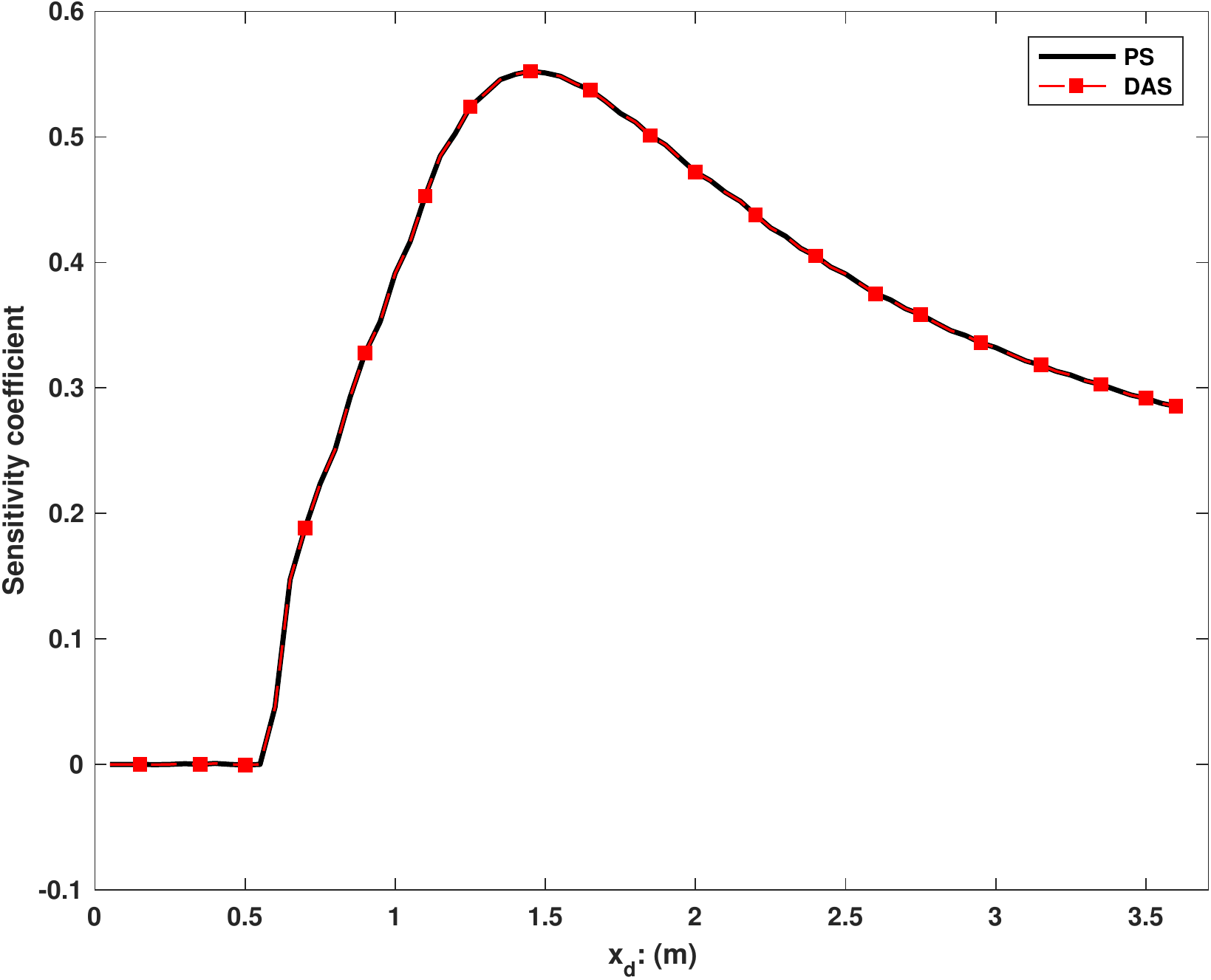}
		\caption{$q = \alpha_g, \omega = Q$}
	\end{subfigure}%
	~
	\begin{subfigure}[t]{0.32\textwidth}
		\centering
		\includegraphics[width=\textwidth, height=0.8\textwidth]{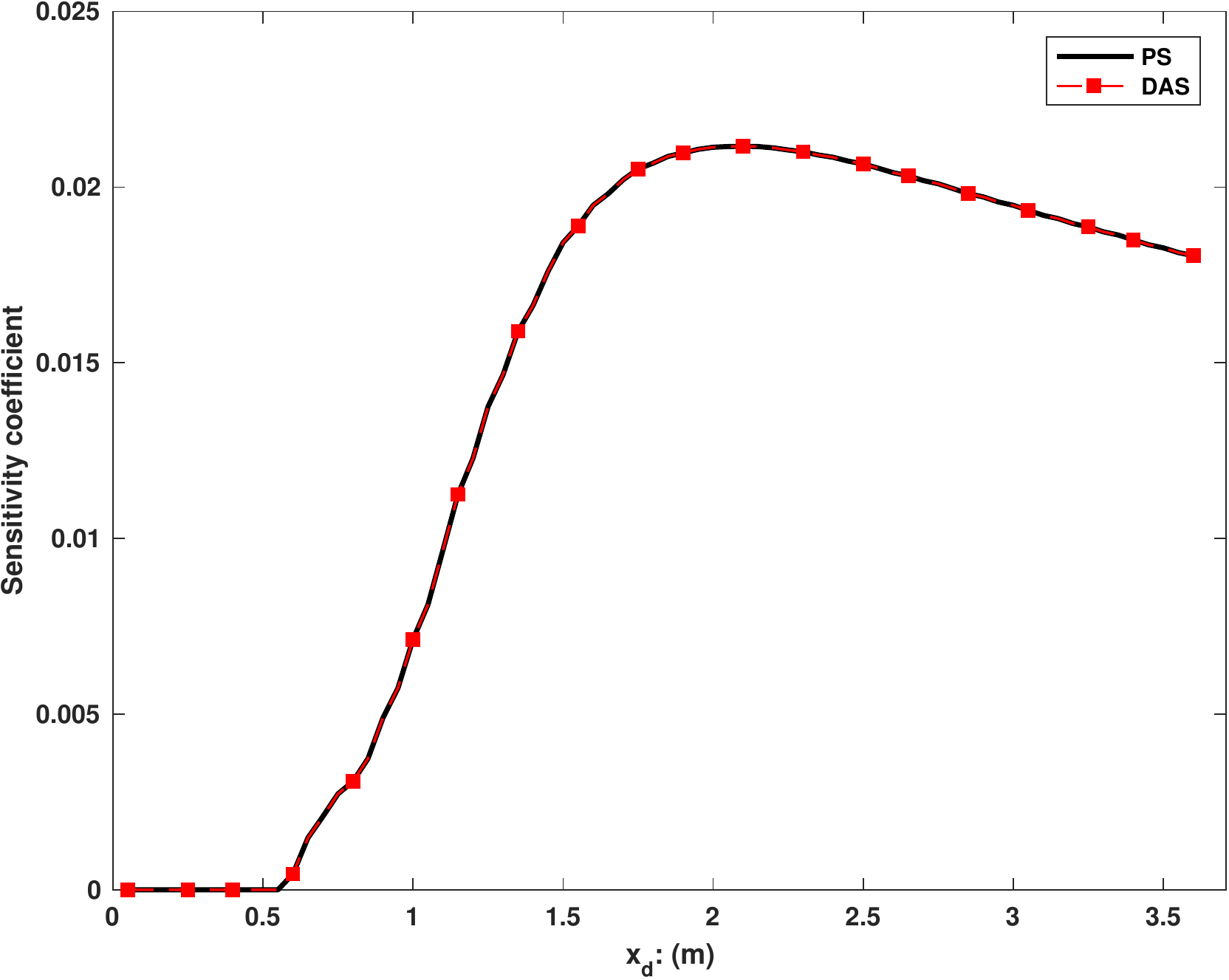}
		\caption{$q = \alpha_g, \omega = f_{i}$}
	\end{subfigure}%
	~
	\begin{subfigure}[t]{0.32\textwidth}
		\centering
		\includegraphics[width=\textwidth, height=0.8\textwidth]{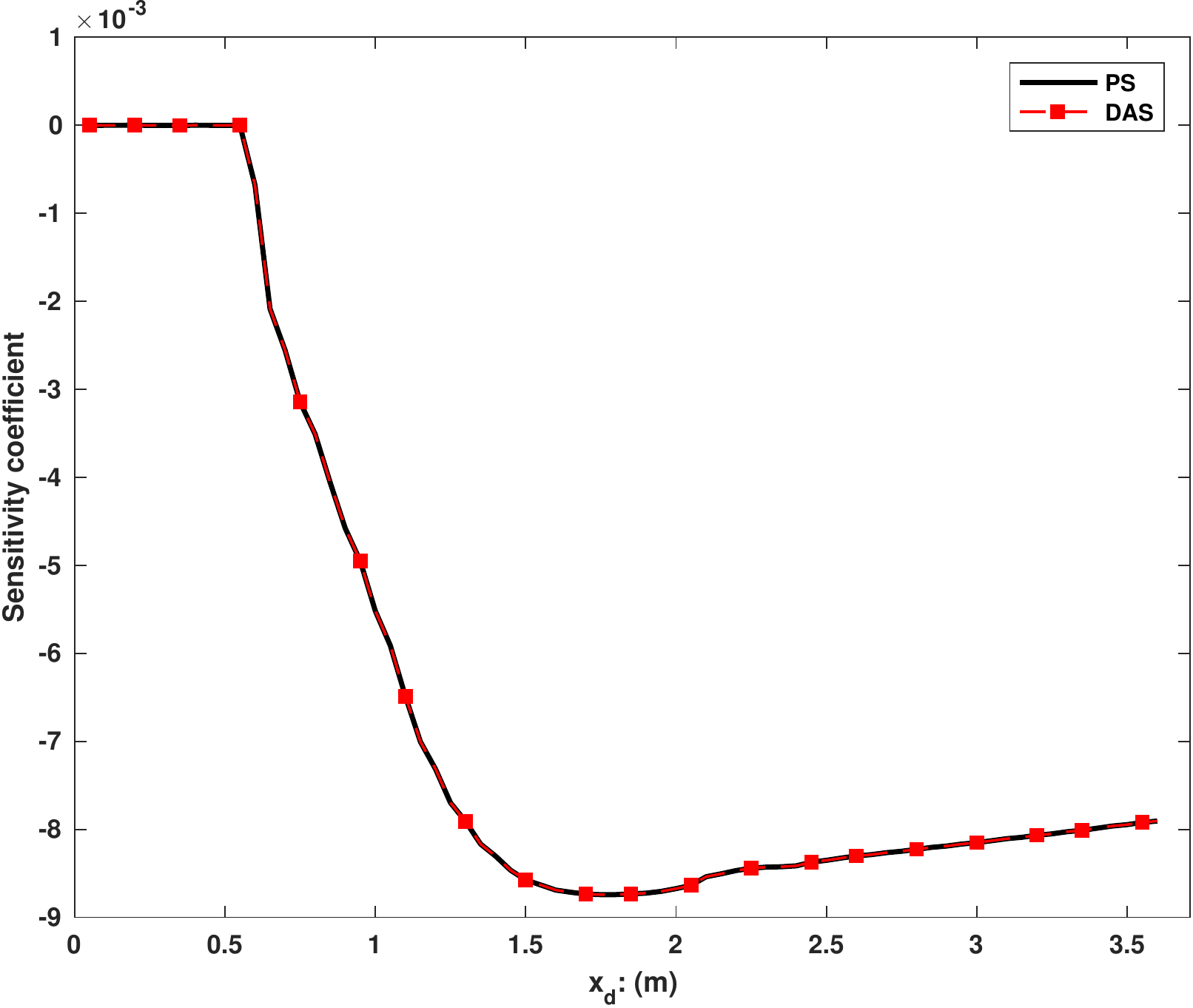}
		\caption{$q = \alpha_g, \omega = f_{wl}$}
	\end{subfigure}%
	\caption{Comparison of sensitivity coefficients from discrete adjoint method (DAS) and perturbation method (PS) for BFBT assembly 4 case 22 at steady-state: continued. $\mathrm{N} = 192$. } \label{SS-BFBT-A4C25-Sensitivity-b}
\end{figure}

\begin{figure}[!htbp]
	\centering
	\begin{subfigure}[t]{0.32\textwidth}
		\centering
		\includegraphics[width=\textwidth, height=0.8\textwidth]{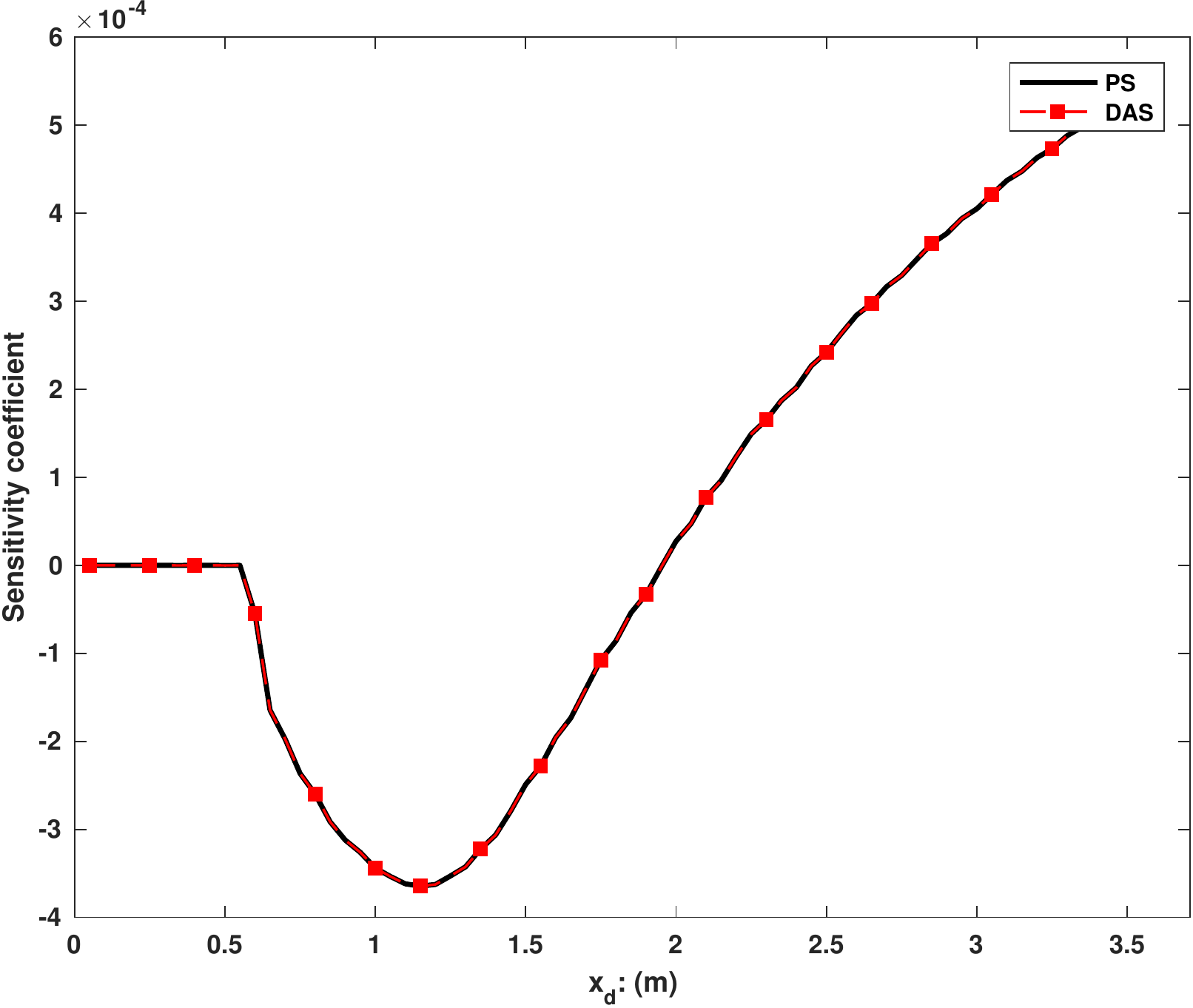}
		\caption{$q = \alpha_g, \omega = f_{wg}$}
	\end{subfigure}%
	~
	\begin{subfigure}[t]{0.32\textwidth}
		\centering
		\includegraphics[width=\textwidth, height=0.8\textwidth]{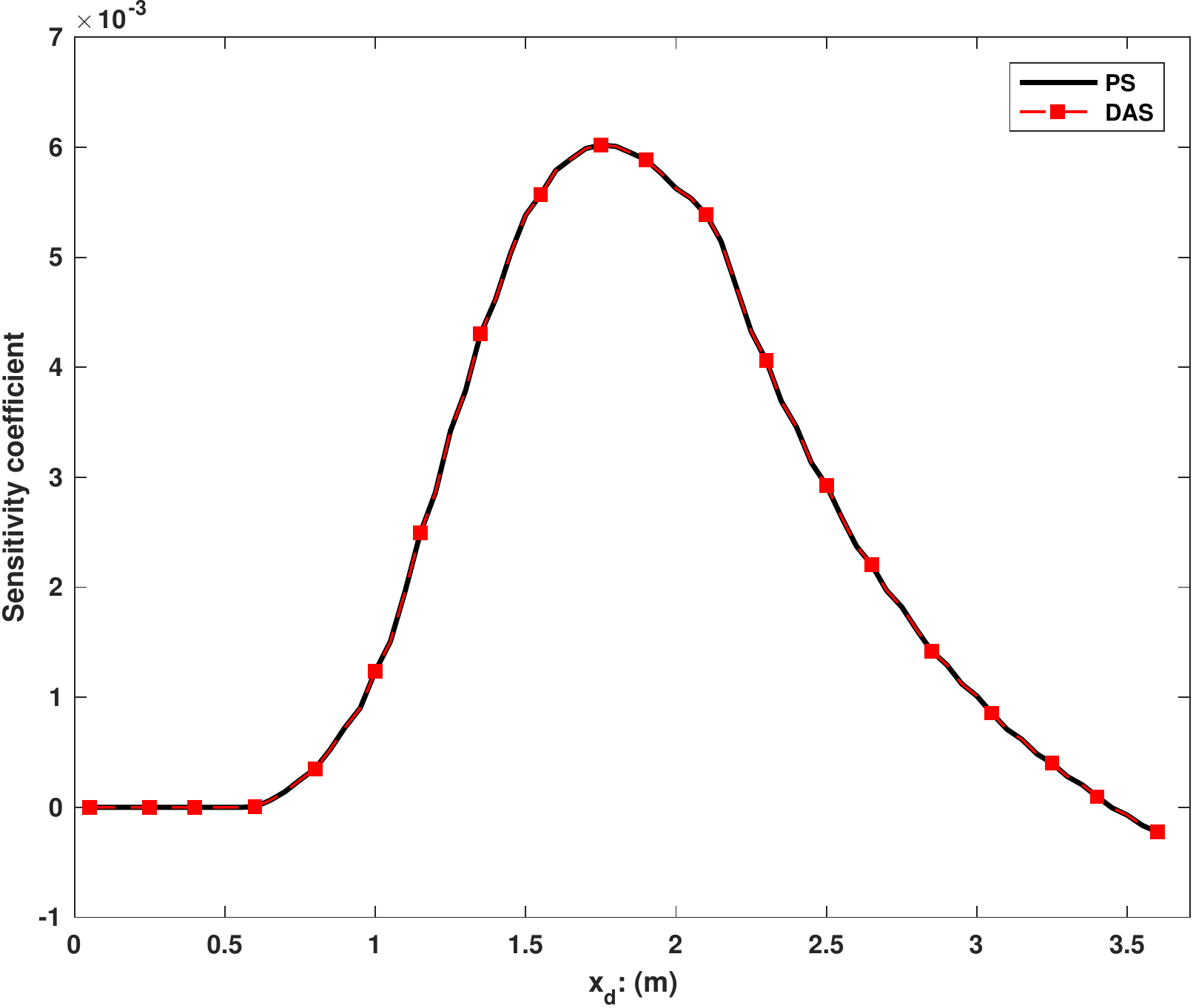}
		\caption{$q = \alpha_g, \omega = H_{il}$}
	\end{subfigure}%
	~
	\begin{subfigure}[t]{0.32\textwidth}
		\centering
		\includegraphics[width=\textwidth, height=0.8\textwidth]{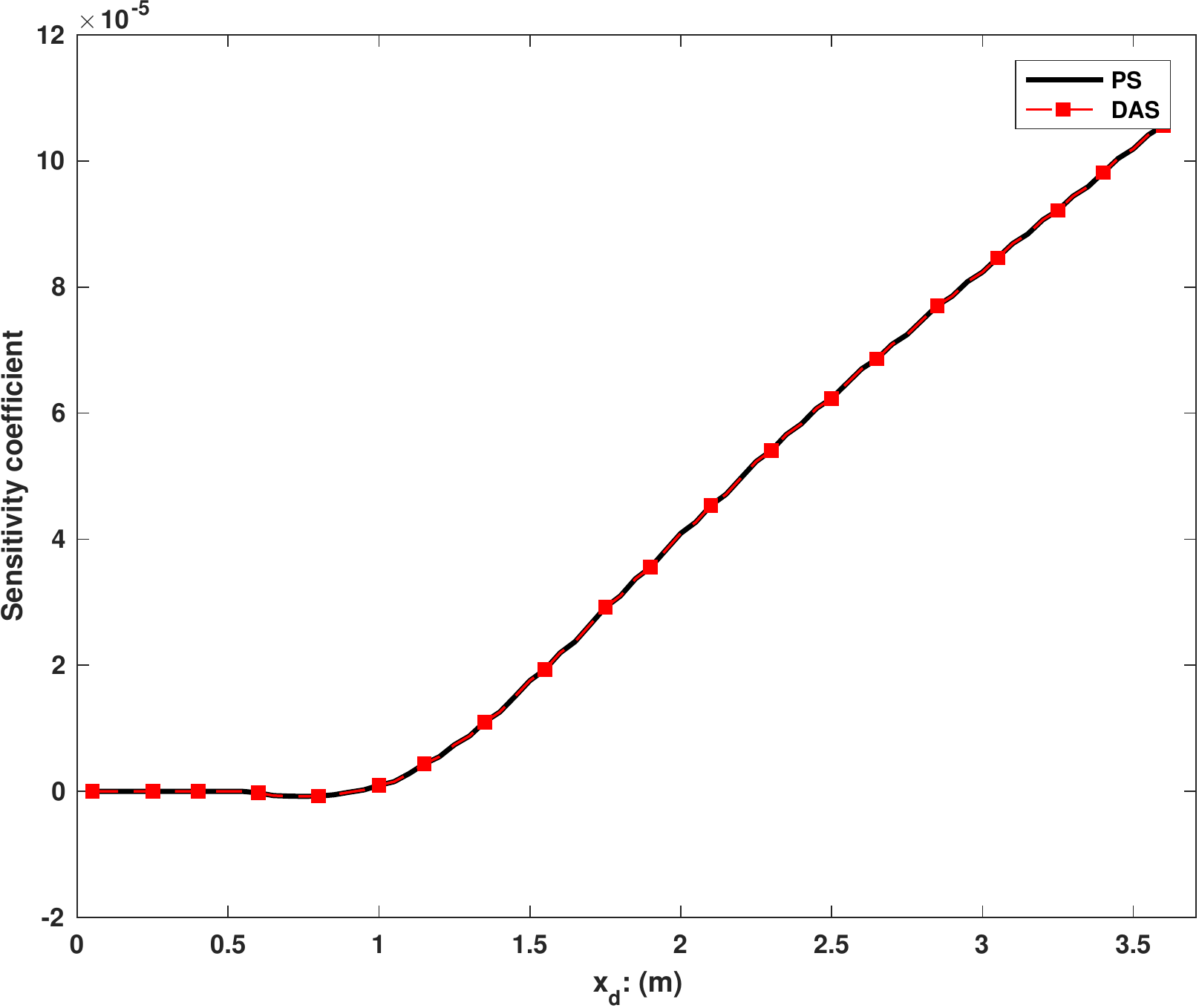}
		\caption{$q = \alpha_g, \omega = H_{ig}$}
	\end{subfigure}%
	\caption{Comparison of sensitivity coefficients from discrete adjoint method (DAS) and perturbation method (PS) for BFBT assembly 4 case 22 at steady-state: continued $\mathrm{N} = 192$. } \label{SS-BFBT-A4C25-Sensitivity-c}
\end{figure}

For this test, responses are taken to be the void fraction at different locations. \refFig{SS-BFBT-A4C25-Sensitivity-a}, \refFig{SS-BFBT-A4C25-Sensitivity-b}, and \refFig{SS-BFBT-A4C25-Sensitivity-c} show the sensitivity coefficients for void fraction to the 9 input parameters. It is seen that the sensitivities from the adjoint method (DAS) match that from the perturbation method (PS). The sensitivity coefficients are reasonable. The behavior of void fraction to these input parameters are summarized as:
\begin{itemize}
	\item In general, the void fraction at the subcooled boiling region ($x_d < 2.0$ m) is more sensitive to the input parameters than the void fraction in the saturated region. The void fraction in the single-phase region ($x_d < 0.55$) does not depend on the input parameters. 
	\item \refFig{SS-BFBT-A4C25-Sensitivity-a}(a): outlet pressure ($p_{outlet}$). Increasing the outlet pressure tends to decrease the void fraction, because the saturation temperature increases with pressure. However, the sensitivity to outlet pressure is small noting that the value in \refFig{SS-BFBT-A4C25-Sensitivity-a}(a) is normalized by the nominal value of the outlet pressure (i.e. $3.943 \times 10^{6}$). 
	\item \refFig{SS-BFBT-A4C25-Sensitivity-a}(b): inlet liquid temperature ($T_{l, inlet}$). Increasing the inlet liquid temperature tends to increase the void fraction, because the subcooling level is increased. The effect of inlet liquid temperature on void fraction is more important in the subcooled boiling region than in the saturation boiling region. 
	\item \refFig{SS-BFBT-A4C25-Sensitivity-a}(c): inlet liquid velocity ($u_{l, inlet}$). Increasing the inlet liquid velocity tends to decrease the void fraction, because the total mass flux is increased but the heating power does not change.
	\item \refFig{SS-BFBT-A4C25-Sensitivity-b}(a): wall to liquid heat flux ($Q_{wl}$). Increasing the wall to liquid flux tends to increase the void fraction, as it is equivalent to increasing the heating power to the rod bundle. 
	\item \refFig{SS-BFBT-A4C25-Sensitivity-b}(b): interfacial friction ($f_i$). Increasing the interfacial friction tends to decrease the relative velocity and increases the void fraction.
	\item \refFig{SS-BFBT-A4C25-Sensitivity-b}(c), \refFig{SS-BFBT-A4C25-Sensitivity-c}(a), (b), and (c): wall friction ($f_{wl}, f_{wg}$) and interfacial heat transfer coefficients ($H_{il}, H_{ig}$). The effect of these input parameters on the void fraction is negligible.
\end{itemize}
Sensitivity information like this can be obtained for other responses, which is omitted for brevity.

\subsubsection{Test 2: uncertainty propagation}
The adjoint sensitivities can be used for efficient propagation of uncertainties in input parameters to uncertainty in the response. Assuming there is no correlation between different input parameters, the standard deviation of the response can be obtained with 
\begin{equation}\label{uncertainty-prop-finite}
\sigma_R^2 = \sum_{m=1}^{9}\normp{\frac{\mathrm{d} R}{\mathrm{d}\omega_{m}}\sigma_{m}}^2 
\end{equation}
where $\sigma_m$ is the standard deviation of the $m^{\mathrm{th}}$ input parameter and $\sigma_R$ is the standard deviation of the response. Note that \refEq{uncertainty-prop-finite} is based on a linear relation between the input parameter and the response.

A Monte Carlo method can be used for uncertainty propagation purposes since there are not many input parameters. The Monte Carlo method can be used to verify the uncertainties obtained with the adjoint method using \refEq{uncertainty-prop-finite}. For most of the input parameters, the uncertainty information (or the PDF) is unavailable. The common practice is to use an ad-hoc distribution that is based on expert judgments. Another promising method is the so-called inverse uncertainty quantification method \cite{hu2016inverse, wu2017inverse}, which quantifies the input uncertainty based on experiment data and code prediction. \refTab{bfbt-monte-carlo-distribution} lists the selected probability distribution function (PDF) of these 9 input parameters. The standard deviation for the outlet pressure, inlet liquid velocity, inlet liquid velocity, and the wall to liquid heat flux (heating power) are obtained from the BFBT benchmark \cite{bfbtV1}. Ad-hoc PDFs are selected for the other 5 physical model parameters, which does not affect the generality of applying the adjoint method to the uncertainty propagation.

\begin{table}[!htb]
	\caption{Probability distribution of boundary conditions and physical model parameters}\label{bfbt-monte-carlo-distribution}
	\centering
	\begin{tabular}{lllll}
		\hline
		Parameter & Parameter name & PDF & Mean & Stand. Dev. \\
		\hline
		$p_{oultlet}$ & Outlet pressure & Normal & Nominal & 1 \%  \\
		$T_{l, inlet}$ & Inlet liquid temperature & Normal & Nominal  & 1.5 K \\
		$u_{l, inlet}$ & Inlet liquid velocity & Normal &Nominal  & 1.0 \% \\
		$Q_{wl}$     & Wall-to-liquid heat flux & Normal & Nominal & 1.5\% \\
		$f_{i}$ & Interfacial friction & Normal & Nominal & 10 \% \\
		$f_{wl}$ & Wall-to-liquid friction & Normal & Nominal & 10 \% \\
		$f_{wg}$ & Wall-to-gas friction & Normal & Nominal & 10 \% \\
		$H_{il}$ & Interface-to-liquid heat transfer coefficient & Normal & Nominal & 10 \% \\
		$H_{ig}$ & Interface-to-gas heat transfer coefficient & Normal & Nominal & 10 \% \\
		\hline
	\end{tabular}
\end{table}

\begin{figure}[!htbp]
	\centering
	\begin{subfigure}[t]{0.45\textwidth}
		\centering
		\includegraphics[width=\textwidth, height=0.8\textwidth]{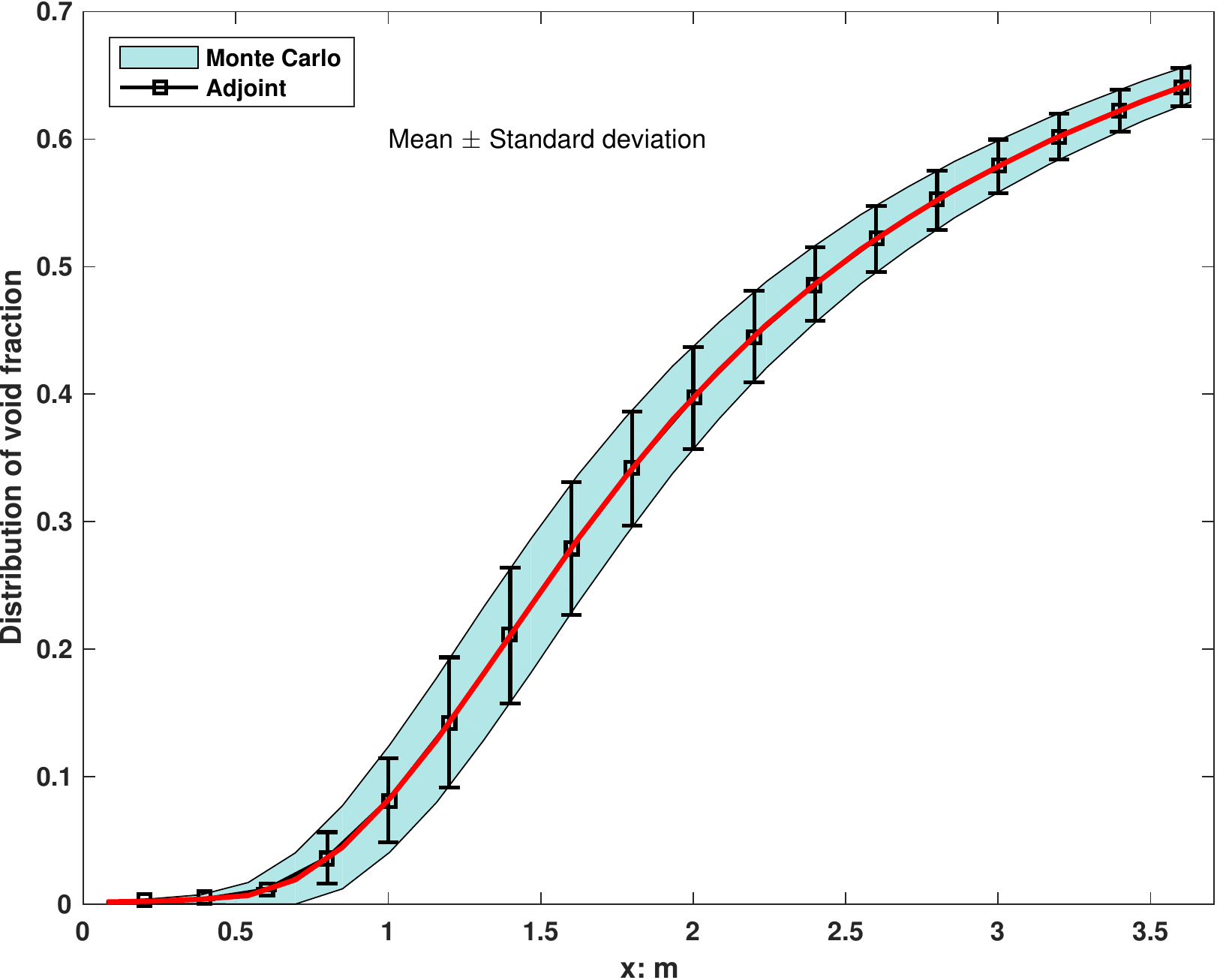}
		\caption{Mean $\pm$ Standard deviation}
	\end{subfigure}%
	~
	\begin{subfigure}[t]{0.45\textwidth}
		\centering
		\includegraphics[width=\textwidth, height=0.8\textwidth]{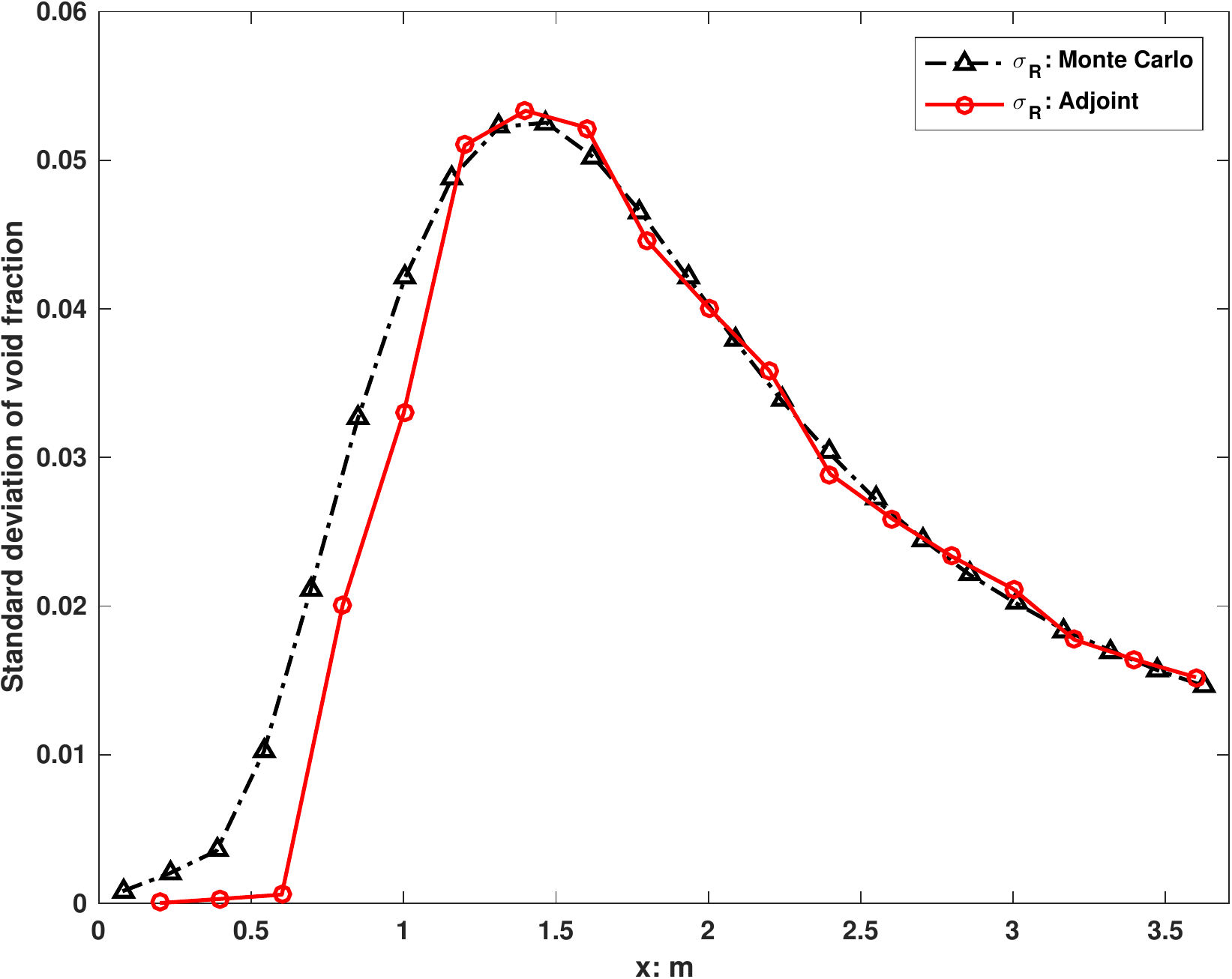}
		\caption{Standard deviation}
	\end{subfigure}%
	\caption{Uncertainty in void fraction calculated with the adjoint and Monte Carlo method. }\label{uq-mc-vs-adjoint}
\end{figure}

The uncertainty in void fraction is calculated by the adjoint and Monte Carlo method (averaged over 200 runs) for comparison. \refFig{uq-mc-vs-adjoint} shows the total uncertainty in the void fraction obtained with both methods. Standard deviation from both methods is also plotted for comparison. In the saturation boiling region, the total uncertainty from the adjoint method matches the uncertainty from the Monte Carlo method very well. However, in the single-phase and subcooled boiling region, there is large difference in the single-phase and subcooled region. This difference is due to the highly nonlinear relation between the boundary conditions and single-to-two-phase transition. The variation in the boundary conditions affects significantly the start point of the subcooled boiling, which is reflected by the Monte Carlo samples but not reflected by the adjoint sensitivities. 

In addition to the total uncertainty, the adjoint method also gives the contribution of each parameter.  \refTab{bfbt-adjoint-component-contribution} shows the uncertainty contribution from the 9 input parameters to the void fraction at 4 measurement locations. For the distribution shown in \refTab{bfbt-monte-carlo-distribution}, it is seen that the inlet liquid temperature contributes the most part. However, these values are not important for practical purposes since only ad-hoc distributions are used. Note that the Monte Carlo method can also give the contribution of each parameter but in the cost of many more runs. 

\begin{table}[!htb]
	\caption{Contribution of input parameters to the uncertainty in void fraction at 4 measurement locations. }\label{bfbt-adjoint-component-contribution}
	\centering
	\begin{tabular}{l|cc|cc|cc|cc}
		\hline
	Location	     & \multicolumn{2}{|c|}{DEN 3} & \multicolumn{2}{c|}{DEN 2}  & \multicolumn{2}{c|}{DEN 1}  & \multicolumn{2}{c}{CT} \\
	\hline
	Parameter      &  $\sigma^2$ & Ratio: \% & $\sigma^2$ & Ratio: \%  &  $\sigma^2$ &  Ratio: \%  & $\sigma^2$ &  Ratio: \%  \\
		\hline
	$p_{oultlet}$ &	2.28E-08	&	6.97	&	3.56E-04	&	14.89	&	9.41E-05	&	17.30	      &	3.88E-05	&	18.02	\\
	$T_{l, inlet}$&	2.96E-07	&	90.41	&	1.94E-03	&	81.30	&	4.02E-04	&	73.86	      &	1.46E-04	&	67.73	\\
	$u_{l, inlet}$&	1.63E-09	&	0.50	&	2.57E-05	&	1.08	&	1.21E-05	&	2.22	      &	7.32E-06	&	3.40	\\
	$Q$           &	2.60E-10	&	0.08	&	5.99E-05	&	2.51	&	3.07E-05	&	5.64	      &	1.92E-05	&	8.91	\\
	$f_{i}$       &	6.47E-09	&	1.98	&	4.21E-06	&	0.18	&	4.34E-06	&	0.80	      &	3.38E-06	&	1.57	\\
	$f_{wl}$      &	1.92E-10	&	0.06	&	7.91E-07	&	0.03	&	8.67E-07	&	0.16	      &	8.16E-07	&	0.38	\\
	$f_{wg}$      &	3.16E-14	&	$<0.01$	&	3.28E-10	&	$<0.01$	&	3.01E-10	&	$<0.01$	      &	1.19E-09	&	$<0.01$	\\
	$H_{il}$      &	2.69E-13	&	$<0.01$	&	3.95E-07	&	0.02	&	1.08E-07	&	0.02	      &	3.25E-09	&	$<0.01$	\\
	$H_{ig}$      &	9.80E-18	&	$<0.01$	&	4.45E-12	&	$<0.01$	&	4.80E-11	&	$<0.01$	      &	1.01E-10	&	$<0.01$	\\
	Total         &	3.27E-07	&	100.00	&	2.39E-03	&	100.00	&	5.44E-04	&	100.00	      &	2.15E-04	&	100.00	\\
		\hline
	\end{tabular}
\end{table}

\subsubsection{Test 3: validation with measurement data}\label{sec-four-p2-t3}
This test is to apply the AdSA and uncertainty propagation to all cases in BFBT assembly 4. The objective is to propagate the uncertainty in the 9 input parameters to the void fraction in 4 measurement locations. There are in total 86 cases in assembly 4. The Monte Carlo method becomes very expensive for this task, because hundreds of forward simulations are required for each case. The adjoint method is very efficient, because only 1 forward simulation is required for each case. The adjoint equation, \refEq{discrete-adjoint-equation-ss}, has to be solved 4 times (1 for each measurement location) for each case, which adds little CPU time to the forward simulation.

\begin{figure}[!htbp]
	\centering
	\includegraphics[width=0.5\textwidth]{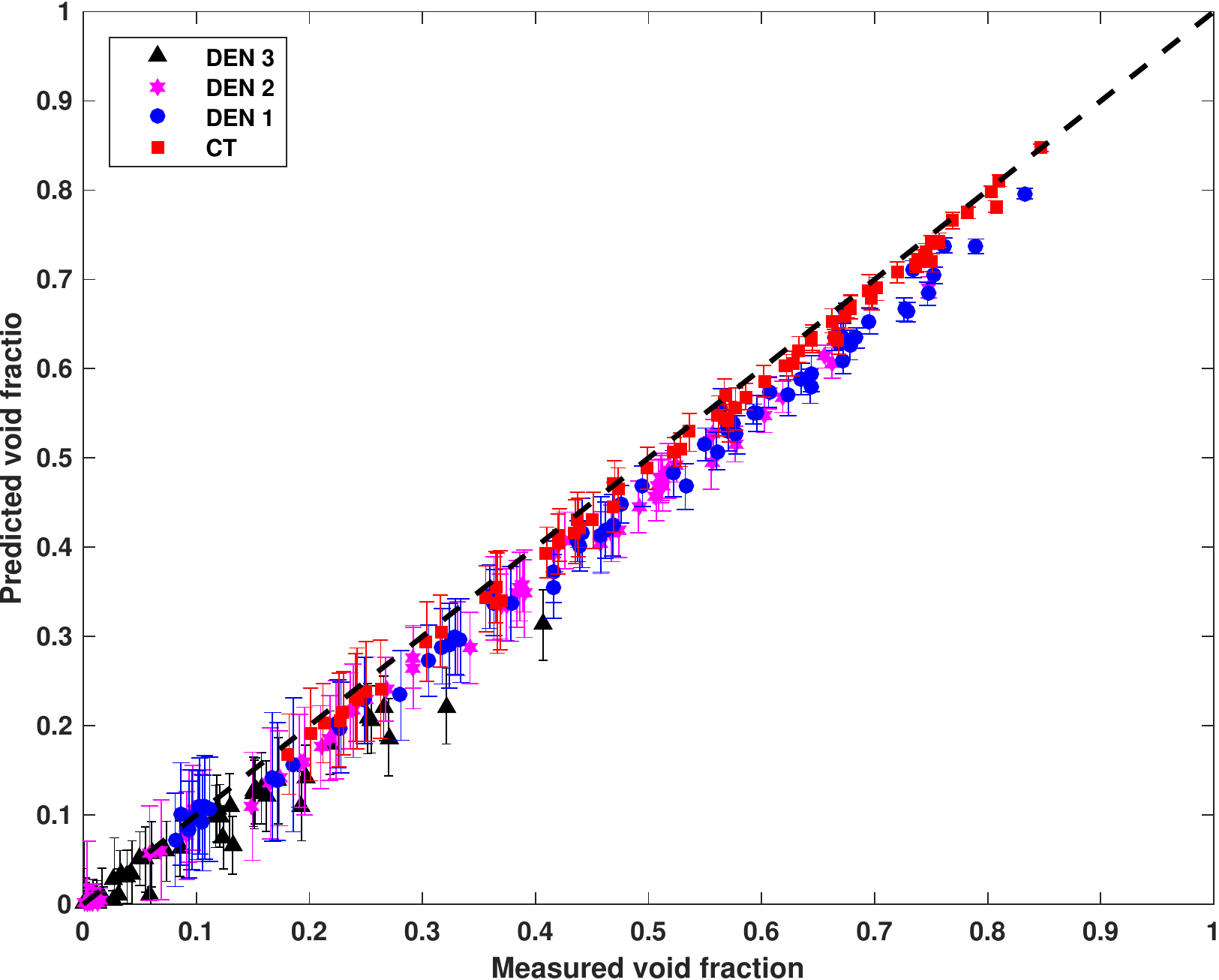}
	
	\caption{Solution of BFBT assembly 4 at steady-state. Nominal value and the standard deviation are plotted. }
	\label{bfbt-assembly-4}
\end{figure}

In this test, 48 cells are used in the simulations. For simplicity, the PDF of 9 input parameters are the same as shown in \refTab{bfbt-monte-carlo-distribution}. The simulation result is shown in \refFig{bfbt-assembly-4}, which includes not only the nominal prediction but also the uncertainty propagated from the 9 input parameters. Because the adjoint sensitivities are available, whenever the PDFs of input parameters are updated, this information can be immediately propagated to the void fraction without new forward simulations. The efficiency of the adjoint method can be best shown by the computational cost. It takes about 5 minutes with a single-core laptop to obtain the result shown in  \refFig{bfbt-assembly-4}.

\section{Conclusion}\label{sec-five}
In this paper, an AdSA framework is developed and verified for sensitivity analysis in steady-state two-phase flow simulations.  The framework is based on the discrete adjoint method and a new implicit forward solver. Numerical tests with the faucet flow problem and the BFBT benchmark verify the adjoint SA framework.  Adjoint sensitivities are shown to match analytical sensitivities very well in the faucet flow problem. The adjoint method is used to propagate uncertainty in input parameters to the void fraction in the BFBT benchmark test. The uncertainty propagation with the adjoint method is verified with the Monte Carlo method and is shown to be efficient. The key features of this adjoint method are:
\begin{itemize}
	\item The method is based on the global residual vector of the forward solver. Extending this method to other forward solvers is straightforward, especially for residual-based implicit solvers. 
	\item The method is capable of providing detailed sensitivity information with little additional cost.   
	\item The method is efficient for calculating sensitivities to a large number of input parameters. Applying this method to uncertainty propagation saves significant amount of forward simulations. Contribution of each parameter to the total uncertainty is  available without additional computation. 
\end{itemize}

The current AdSA framework is limited to steady-state two-phase flow simulations. Extending this framework to transient two-phase flow simulations is an important future work. 

\section*{Acknowledgment}
The authors would like to thank the anonymous reviewers for their detailed and valuable comments to improve the quality of the paper.

\appendix
\section{Approximate eigenvalues/eigenvectors}
The Jacobian matrix $\mat{A}_c$ is
\begin{equation}\label{Eq-App-17}
\mat{A}_c = \scalebox{.9}{$\begin{pmatrix}
	0 & 1 & 0 & 0 & 0 & 0 \\
	-u_l^2+\beta_l c_{l}^{h} & 2u_l-\beta_l c_{l}^{u} & \beta_l c_{l}^{1} & \sigma_l c_{g}^{h} & -\sigma_l c_{g}^{u} & \sigma_l c_{g}^{1} \\
	-u_l H_l +  u_l\beta_l c_{l}^{h} & H_l -u_l\beta_l c_{l}^{u} & u_l+u_l\beta_l c_{l}^{1} & \sigma_l u_l c_{g}^{h} & -\sigma_l u_l c_{g}^{u} & \sigma_l u_l c_{g}^{1} \\
	0 & 0 & 0 & 0 & 1 & 0 \\
	\sigma_g c_{l}^{h} & -\sigma_g c_{l}^{u} & \sigma_g c_{l}^{1} & -u_g^2+\beta_g c_{g}^{h} & 2u_g-\beta_g c_{g}^{u} & \beta_g c_{g}^{1} \\
	\sigma_g u_g c_{l}^{h} & -\sigma_g u_g c_{l}^{u} & \sigma_g u_g c_{l}^{1} & -u_g H_g + u_g\beta_g c_{g}^{h} & H_g -u_g\beta_g c_{g}^{u} & u_g+u_g\beta_g c_{g}^{1} \\
	\end{pmatrix}$}
\end{equation}
where
\begin{subequations}\begin{align}
	c_l^h \equiv a_l^2 + \normp{\gamma_l -1}\normp{u_l^2 - H_l} &;\quad c_g^h \equiv a_g^2 + \normp{\gamma_g -1}\normp{u_g^2 - H_g}\\
	c_l^u \equiv \normp{\gamma_l -1}u_l &;\quad c_g^u \equiv \normp{\gamma_g -1}u_g \\
	c_l^1 \equiv \gamma_l -1 &;\quad c_g^1 \equiv \gamma_g -1 \\
	\beta_{l} \equiv \frac{1+\alpha_l\varepsilon_g}{1 + \alpha_g\varepsilon_l + \alpha_l\varepsilon_g} &;\quad \beta_{g} \equiv \frac{1+\alpha_g\varepsilon_l}{1 + \alpha_g\varepsilon_l + \alpha_l\varepsilon_g} \\
	\sigma_{l} \equiv \frac{\alpha_l \varepsilon_l}{1 + \alpha_g\varepsilon_l + \alpha_l\varepsilon_g} &;\quad \sigma_{g} \equiv \frac{\alpha_g \varepsilon_g}{1 + \alpha_g\varepsilon_l + \alpha_l\varepsilon_g}
	\end{align}\end{subequations}
\begin{equation}\label{Eq-A.15}
a_k^2 \equiv \frac{1}{\normp{\frac{\partial \rho_k}{\partial p}}_{h_k} + \frac{1}{\rho_k}\normp{\frac{\partial \rho_k}{\partial h_k}}_{p}}, \quad \gamma_k \equiv \frac{\normp{\frac{\partial \rho_k}{\partial p}}_{h_k}}{\normp{\frac{\partial \rho_k}{\partial p}}_{h_k} + \frac{1}{\rho_k}\normp{\frac{\partial \rho_k}{\partial h_k}}_{p}},\quad \varepsilon_k = \frac{\rho_k a_k^2 - \gamma_k p}{p}
\end{equation}
The approximate eigenvalues and right eigenvectors are
\begin{subequations}\label{theory:refEq-B2}\begin{align}
	\lambda_{c,1} \approx u_l - \sqrt{\beta_l}a_l; \lambda_{c,2} &= u_l ; \lambda_{c,3} \approx u_l + \sqrt{\beta_l}a_l \\
	\lambda_{c,4} \approx u_g - \sqrt{\beta_g}a_g; \lambda_{c,5} &= u_g ; \lambda_{c,6} \approx u_g + \sqrt{\beta_g}a_g
	\end{align}\end{subequations}
\begin{equation}\label{theory:refEq-B3}\begin{split}
\vect{K}_{1,c} \approx \begin{pmatrix}
1 \\
u_l -\sqrt{\beta_l}a_l \\
H_l -\sqrt{\beta_l}a_l u_l \\
0 \\
0 \\
0 \\
\end{pmatrix},
\vect{K}_{2,c} &\approx  \begin{pmatrix}
1 \\
u_l \\
H_l - \gamma_l^{*}a_l^2 \\
0 \\
0 \\
0 \\
\end{pmatrix},
\vect{K}_{3,c} \approx  \begin{pmatrix}
1 \\
u_l +\sqrt{\beta_l}a_l \\
H_l +\sqrt{\beta_l}a_l u_l \\
0 \\
0 \\
0 \\
\end{pmatrix}\\
\vect{K}_{4,c} \approx \begin{pmatrix}
q_4\\
q_4\lambda_{c,4} \\
q_4\normb{H_l - u_l^2 + u_l\lambda_{c,4}}\\
1 \\
u_g -\sqrt{\beta_g}a_g \\
H_g -\sqrt{\beta_g}a_g u_g \\
\end{pmatrix},
\vect{K}_{5,c} &\approx  \begin{pmatrix}
0\\
0\\
0\\
1 \\
u_g \\
H_g - \gamma_g^{*}a_g^2 \\
\end{pmatrix},
\vect{K}_{6,c} \approx  \begin{pmatrix}
q_6\\
q_6\lambda_{c,6}\\
q_6\normb{H_l - u_l^2 + u_l\lambda_{c,6}}\\
1 \\
u_g +\sqrt{\beta_g}a_g \\
H_g +\sqrt{\beta_g}a_g u_g \\
\end{pmatrix}
\end{split}\end{equation}
where $\gamma_l^{*}=1/\normp{\gamma_l-1}$ and $\gamma_g^{*}=1/\normp{\gamma_g-1}$. $q_4$ and $q_6$ are two auxiliary variables defined as
\begin{equation}
q_4 \equiv \frac{\sigma_l a_g^2}{\normp{\lambda_{c,4}-\lambda_{c,1}}\normp{\lambda_{c,4}-\lambda_{c,3}}}; \quad
q_6 \equiv \frac{\sigma_l a_g^2}{\normp{\lambda_{c,6}-\lambda_{c,1}}\normp{\lambda_{c,6}-\lambda_{c,3}}}
\end{equation}


\end{document}